\documentclass{article}%
\usepackage{graphicx}
\usepackage{amsmath}
\usepackage{amsfonts}
\usepackage{amssymb}%
\providecommand{\U}[1]{\protect\rule{.1in}{.1in}}

\begin{document}

\title{Pascual Jordan's legacy and the ongoing research in quantum field theory\\{\small dedicated to my teacher and role model: Rudolf Haag}\\{\small appeared in "Historical Perspectives on Contemporary Physics" Eur.
Phys. \textbf{35}, J.H. 377 (2010)}}
\author{Bert Schroer\\present address: CBPF, Rua Dr. Xavier Sigaud 150, \\22290-180 Rio de Janeiro, Brazil\\email schroer@cbpf.br\\permanent address: Institut f\"{u}r Theoretische Physik\\FU-Berlin, Arnimallee 14, 14195 Berlin, Germany}
\date{December 2009}
\maketitle

\begin{abstract}
Pascual Jordan's path-breaking role as the protagonist of quantum field theory
(QFT) is recalled and his friendly dispute with Dirac's particle-based
relativistic quantum theory is presented as the start of the field-particle
conundrum which, though in modified form, persists up to this date. Jordan had
an intuitive understanding that the existence of a causal propagation with
finite propagation speed in a quantum theory led to radically different
physical phenomena than those of QM. The conceptional-mathematical
understanding for such an approach began to emerge only 30 years later. The
strongest link between Jordan's view of QFT and modern "local quantum physics"
is the central role of causal locality as the defining principle of QFT as
opposed to the Born localization in QM.

The issue of causal localization is also the arena where misunderstandings led
to a serious derailment of large part of particle theory e.g. the
misinterpretation of an infinite component pointlike field resulting from the
quantization of the Nambu-Goto Lagrangian as a spacetime quantum string.

The new concept of modular localization, which replaces Jordan's causal
locality, is especially important to overcome the imperfections of gauge
theories for which Jordan was the first to note nonlocal aspects of
\textit{physical} (not Lagrangian) charged fields.

Two interesting subjects in which Jordan was far ahead of his contemporaries
will be presented in two separate sections.

\end{abstract}
\tableofcontents

\section{Preface}

Some years ago J\"{u}rgen Ehlers (1929-2008), Jordan's first postwar PhD
student and the founding director of the AEI in Golm, asked me to use my
expertise of quantum field theory (QFT) to check out some papers by Pascual
Jordan with a seemingly somewhat inaccessible content which appeared to have
been left out in existing biographies and accounts of his scientific legacy.
It took me several years to fully appreciate their content and to understand
that they were early harbingers of subtle quantum field theoretic properties
which, in a somewhat different context and a broader setting, had their
comeback in the QFT of the 60s and 70s; in the case of the 1924/25
Einstein-Jordan conundrum one even needs the recent insight about the
connection between modular localization, vacuum polarization at the
localization boundary and thermal aspects from the restriction of the vacuum
state to localized algebras; in other words the full understanding of the QFT
model with which everything begun requires the knowledge of the most advanced
notions of local quantum physics.

I included some of these results in a talk with a large biographical part
which I presented under the title: "Pascual Jordan, biographical notes, his
contributions to quantum mechanics and his role as a protagonist of quantum
field theory" at a conference dedicated to the memory of Pascual Jordan 2005
in Mainz, Germany whose written account can be found online \cite{Mainz}. Here
the biographical material is left out and the presentation of Jordan's QFT
contribution will be extended.

As a quantum field theorist I am intrigued by the modernity of Jordan's view
of the subject. In his contribution in the famous 1925 Dreimaennerarbeit,
which is nowadays considered as the cradle of QFT, Einstein's at that time
still controversial photon ("Nadelstrahlung") in terms of thermal fluctuations
in a subvolume was carried forward into the new quantum theory with the help
of a subinterval in a two-dimensional (conformal) free quantum "photon" field.
The problem of subinterval fluctuation was treated in terms of quantum
mechanical infinitely many oscillators which, as we know nowadays, does not
adaequately describe the holistic quantum field theoretic fluctuation problem
related to localized subvolumes. But already in his 1928 Habilitationsschrift
and in his famous paper with Pauli, the property of causal locality as the
characterizing principle of QFT took the center stage and the fundamental
difference between QFT and Dirac's idea of a relativistic quantum mechanics
(QM) begun to be appreciated. My interest to write a paper about to what
extend modern QFT has vindicated Jordan's ideas arose in this context.

What started as a friendly turn to my highly regarded senior colleague (and
mentor during my brief participation in Jordan's seminar around 1956), the
late J\"{u}rgen Ehlers, changed into surprise when I looked at some of the
insufficiently understood publications of Jordan. It became gradually clear to
me that these papers constitute early flare-ups of the intrinsic conceptual
logic of QFT whose correct mathematical understanding and physical
interpretation was a task beyond the level of QFT at the time of Jordan: even
nowadays those issues would hardly be accessible without foundational
knowledge about local quantum physics; this explains in a way why those
contributions have been left out in presentations of Jordan's oeuvre by other authors.

Already at that time of the Mainz conference I thought that it would be
interesting to present some of Jordan's work written in the years 1930-38
which was lost in the maelstrom of history\footnote{These were the years in
which according to a remark by Peter Bergmann a publication in Zeitschrift
fuer Physik "was like a first class burial".} in more details, in order to
reveal its startling modernity. Though as a passionate champion of research on
the frontier of QFT, I did not want to subject myself to the time-consuming
rules of documenting historical events, but I find the task of creating
conceptual and philosophical connections to ideas which, either as a result of
world war II, or through other sociological reasons, were lost or took a
different turn, quite challenging. In fact there is no better critical view of
ongoing particle physics than that obtained from comparing the present state
of affairs with what the protagonist of QFT expected from his brainchild.

With the launch of the new Springer journal EPJ H, dedicated to the historical
illumination of actual research problems, it is now possible to recreate some
of these lost ideas and demonstrate how deep thoughts occasionally flare up
and disappear for some time before they firmly and permanently assert
themselves in the actual\textbf{ }research\textbf{. }Ideas which have no
continuous link to present day QFT, or which led to profound errors are
presumably not of much interest to historians, but to the extend that they
have left traces which are important for the future of particle theory, they
should be presented next to the successes. The present paper is strictly
limited to the subject in the title and in the abstract. For biographical and
historical details I refer to a previous paper \cite{Bert}. A rather complete
bibliographical list has been compiled in \cite{Bei}.

\section{The first phase of Quantum Field Theory}

Quantum field theory (QFT) is often referred to as relativistic quantum
mechanics (QM), but this characterization has no convincing conceptual or
historical justification. Relativistic QM exists, but the conceptual as well
as computational unfolding of QFT has increased the distance between the two.
The conceptual-mathematical meaning of "relativistic quantum mechanics" (QM)
in the discourse between Jordan and Dirac was somewhat vague, but we know by
now how a consistent relativistic quantum mechanics of interacting particles
looks like, and it has no similarity with interacting QFT besides the fact
that both theories are quantum theories (QT) and carry a unitary
representation of the Poincar\'{e} group (section 3.1).

Historically the birth of field quantization dates back to Jordan's
contribution to the 1925 "Dreim\"{a}nnerarbeit" \cite{B-H-J} which consisted
in the first quantum theoretic derivation of Einstein's fluctuation formula,
thus proving that presence of the wave term and the particle-like (photon)
contribution on which Einstein \cite{Einstein1} based his photon hypothesis
long before the latter was observationally verified in experiments by Compton
as well as by Bothe and Geiger. Hence QFT, which gave rise to many new
problems, originated itself at the same time as QM in the context of a
solution of a problem posed by Einstein. But even at the time of the
publication of the BHJ paper, as well as during all the time since Einstein's
famous 1905 "photon" paper (confirmed and strengthened in his 1917 work
\cite{Einstein1}), there remained a certain reservation against Einstein's
indirect pro photon arguments for which his reasoning based on the
fluctuations in a finite volume formula was his most forceful counter argument.

Even Jordan's two coauthors of the Dreim\"{a}nnerarbeit were not entirely
convinced by his arguments which went against the, at that time still rather
popular, theory by Bohr, Kramers and Slater who tried to do treat radiation by
semiclassical methods in the Bohr-Sommerfeld setting i.e. without photons;
these doubts lingered on for several years to come until that theory was
abandoned. However serious doubts about quantum mechanical fluctuation being
able to substitute Einstein's thermal fluctuation and limit the frequency
summation just to avoid divergencies. As Duncan and Janssen recently recalled
\cite{Du-Ja} (and the present author certainly agrees with, despite some
critical later remarks), Jordan's field theoretic model in support of a
quantum theoretical derivation of Einstein's thermal fluctuation formula,
despite some shortcomings, really heralded the beginning of QFT. However the
concepts and methods which are needed in order to show that the restriction of
the pure vacuum state to a subinterval generates an impure state of a thermal
kind (not possible in QM!) which is related to Einstein's statistical
mechanics fluctuation calculation is an insight into QFT which are of more
recent origin \cite{Ei-Jo} (more in section 4). For this task one needed to
understand the thermal aspect \ of modular localization which generalizes the
well-known observations which Hawking and Unruh made in specific cases about
thermal aspects of quantum matter behind or in front of black hole horizons or
behind causal horizons associated with a Rindler wedge region. The modular
localization theory permits to show that these vacuum fluctuation-caused
effects from localization to subregions really explains the thermal behavior
as derived by Einstein from pure stastisical mechanics considerations.

Ironically the strongest argument that this model represents genuine QFT and
not some infinite degrees of freedom second quantized QM (as the work of
Jordan and Klein \cite{J-K}) is the fact that neither Jordan nor anybody
afterwards was able to present a rigorous pure quantum mechanical derivation
about a free field theory without invoking additional assumptions and
plausibility arguments at some places. The restriction of a quantum mechanical
vacuum to a subinterval causes a tensor-factorization of the Hilbert space and
the operator algebra. In QFT this changes radically: there is no
inside-outside tensor factorization\footnote{Hence also the notion of
entanglement losses its meaning. Only if one prevents the localization region
from touching its causal complement by the "splits" procedure one starts to
control Heisenberg's boundary vacuum polarization clouds and returns to a
tensor factorization, but the reduced split vacuum state remains impure and
thermal.} in any state and the vacuum reduced to the interval is highly impure
\cite{interface}. QM can never lead to a impure thermal state merely by
restricting the vacuum to a subinterval; the Born-localization related to
eigenfunctions of the position operator has no thermal or impure aspects,
rather it shares its concepts of entanglement with quantum information theory.
With (uncontrolled) additional assumptions which destroy the holistic
localization property of QFT, QM yields a formula consisting of a wave and a
particle contribution \cite{B-H-J}, but it is not really capable to fully
solve the Einstein-Jordan conundrum, since the explaining quantum theory
should identify the reduced vacuum state as a mixed state with thermal
properties. QFT on the other hand accomplishes this magic. We know about this
thanks to many decades of research in local quantum physics which has led in
particular the recent insights into \textit{modular localization}
\cite{interface}\cite{BMS}\cite{Ei-Jo}.

Quantum mechanical methods of mode decomposition and maintaining finiteness by
occupying certain modes are generally not correct since they destroy causal
localization; there is hardly any bigger conceptual-mathematical contrast as
that between the Born-localization of QM and the modular localization of QFT
\cite{interface}. The latter forces the theory to behave in a much more
holistic way than QM. Even though the discovery of QFT took place within the
same year as QM, the elaboration of the computational and conceptual
instruments to understand and explore it has lasted already for more than 80
years and is still not anywhere near its closure (more remarks in section 4),
whereas the principles of QM were fully understood within less than two
decades after their discovery.

Actually the prelude to this most fascinating episode in the history of
quantum physics started already with Jordan's thesis \cite{thesis} which he
defended and submitted to Zeitschrift f\"{u}r Physik in 1924, i.e. the year
before the appearance of the famous work of Heisenberg on quantum mechanics
(QM) and the Dreim\"{a}nnerarbeit. The topic of his thesis was a quantum
theory of radiation, still without photons, or in the jargon of the pre-photon
times, without Einstein's "Nadelstrahlung" (the result of the photon recoil).
By submitting his thesis for publication with Born's blessings (besides
Einstein and Pauli there were no strong defenders of the photon hypothesis),
Jordan \cite{thesis} set out to openly contradict Einstein \cite{Einstein2} by
showing that the existence of Nadelstrahlung was not necessary in order to
establish thermodynamic equilibrium. He did not have to wait long for an
equally forceful reaction. Einstein conceded that there was nothing wrong in
Jordan's mathematics, but that in a pure wave picture without the (photon)
Nadelstrahlung the calculation of the correct radiation absorption
coefficients cannot be obtained.

This episode was Jordan's figurative road to Damascus as far as his conversion
to the existence of photons was concerned. Even if, as Born and especially
Heisenberg believed, Jordan did not completely solve the fluctuation
conundrum, he presented us with the beginnings of the richest theory of
quantum matter up to date: QFT. By the time he wrote his section in the
Dreimaennerarbeit he had given up the attempt to obtain thermal equilibrium
without photons and became the radical annunciator of a QFT of waves,
including the de Broglie matter waves. He took great pride about his ability
to explain Einsteins fluctuation conundrum, even if his colleagues maintained
some reservations. Being strongly philosophically oriented, he could not
tolerate a wave quantization for light and quantum mechanics for matter, it
had to be a unified setting for both. His intention to draw Einstein towards
this new conceptual setting was however less successful; Einstein had a lot of
praise for Jordan's abilities, but he never subscribed to QFT or QM as the
final quantum matter theory; the philosophical aspects which attracted Jordan
(the probabilistic aspects of quantum theory was no problem for him) were
exactly those which did not find the likings of Einstein, although both held
the universality of physical principles in high esteem.

The friendly competion with his adversary Dirac, who up to the beginning of
the 50s upheld the viewpoint that matter should be described in terms of
infinitely many quantum mechanical oscillators and the wave quantization
should be reserved for light (just as in classical wave/particle theory), was
more substantial because it took place between two individuals who fully
accepted quantum theory; in fact it was of tremendous benefit to the
development of QT, both for the particle as well as the field side. Even those
discoveries of Dirac which later revealed themselves to be inconsistent on a
deeper level\footnote{The inconsistency onely showed up in higher order
perturbation theory involving vacuum polarization. They did not affect the
computation in Heitler's \cite{Heitler} and Wenzel's book, but it is not
possible to formulate renormalization theory in this charge-unsymmetric
setting. The repair of this theory leads to QED.}, as the particle-inspired
hole theory, came together with gems of permanent endurance as the Dirac
equation. The latter was instrumental in Wigner's representation theoretic
analysis \cite{Wig} of relativistic one-particle states, the first completely
intrinsic relativistic QT (without any quantization parallelism).

To be conceptually and mathematically precise on this point of QFT versus QM,
one should add that relativistic QM really exists as a automomous consistent
theory (the inconsistency of Dirac's hole theory did not dispel it); but in
its conceptual-mathematical details it is quite different from what Dirac
thought is should be. The relativistic aspect (representation of the
Poincar\'{e} group) formulated in terms of particles leads to a multiparticle
representation which, in the presence of interaction potentials, cannot be
written down by adding up pair potentials (as in the nonrelativistic QM where
the cluster factorization of the n-particle factorization and the S-matrix are
obtained without any effort). Rather the 3-particle interaction has to be
determined from the two-particle potential by an interplay of relativity and
cluster arguments and not by simply adding up two-particle potentials; this
"cluster" construction has to be inductively implemented.

Since in the step from n to n+1 particles the upholding of the cluster
property demands the introduction of a connected "minimal" n+1-particle
interaction potential which is induced in terms of the previously determined
interaction potentials, there is not intrinsic distinction between elementary
interaction potentials and induced ones. This "direct particle interaction"
(DPI) (i.e. not field mediated) \cite{C-P}, unlike the Schr\"{o}dinger QM,
does not admit a natural "second quantization formalism"\footnote{This is of
course related to the fact that the cluster factorization is not the result of
the additivity of interaction potentials.}, even though its multiparticle
Poincar\'{e} group representations cluster-factorize.

The existence of such relativistic particle theories (called direct particle
interactions (DPI)), which satisfy all the conceptual properties which one can
express \textit{in terms of particles without using interpolating fields}, is
not very interesting by itself; but by creating a sharp contrast to the causal
locality of QFT it his of invaluable help in understanding the subtleties of
the latter. Most of the serious errors committed in the last 5 decades, in
particular the misunderstanding that string theory has something to do with
spacetime localized strings and that matters of localization can be red off
from the S-matrix, result from misconceptions about QM and QFT localization;
in fact all differences between these two quantum theories, which, as
mentioned, do not disappear by making QM relativistic, originate from their
different localization, as will be shown in the next section.

Since interactions (which in QFT would be fixed in terms of local coupling
parameters), need potential functions instead of coupling parameters in order
to be defined, the DPI setting is certainly not "fundamental"\footnote{Nuclear
physicists use the DPI setting to decribe elastic scattering (or the creation
of only a few particles) in intermediate energy regions.}. Its very existence
permits to immediately refute some conjectures concerning the relation between
particles and fields. In particular Weinberg's idea that the unitarity,
Poincar\'{e} invariance and cluster factorization of the S-matrix can only
originate in a QFT setting \cite{cluster} is not correct, it remains even
incorrect if one adds to the list the timelike form of macro-causality
(Stueckelberg's "causal re-scattering"). Of course Weinberg (as well as the
present author) would never consider DPI as a fundamental relativistic QT; but
for showing the sublety of causal localization (which cannot be expressed in
terms of particles and their clustering) it is an ideal surgeons knife which
separates the particle from the field body. In the present context a
rudimentary conceptual knowledge of DPI is valuable because it serves to lift
the field-particle relation, personified by Jordan and Dirac, to a higher
level than it has been discussed in most historically motivated articles. In
contrast to the particle-wave conundrum of the beginnings of QT, which
disappeared with the understanding of the equivalences between different
formulation of QM and the ascend of QFT, the field-particle relation, despite
progress, has not yet been closed.

In fact since the more than 80 years of its existence of QFT it was always
with us, and the attempts to get rid of it by adopting a pure S-matrix view
failed in each attempt, and even nowadays there is a large community which was
hitherto unable to liberate itself from the self-knitted webb of conceptual
confusions. Although there has been great progress, there is as yet no closure
on this issue; in fact it appears as if the decisive nonperturbative
understanding concerning an intrinsic classification, as well as an existence
proof and a construction of models is still to come\footnote{This goal has
been achieved to some degree in d=1+1 (chiral models, factorizing models)
based on modular localization in the setting of operator algebras.}. So a
large part of this presentation of Jordan's QFT uses history as a pretense to
lure the historically educated interested reader into the ongoing "theoretical
laboratory", or in the other direction, to reveal to the professional quantum
field theorist the ideas which marked the beginning of QFT. The aim is always
to get a new perspective about the the relation of the ongoing research and
its history. This is also the best way to acquire some immunity against the
incorrect idea that QFRT arrived at its closure and is on its way of becoming
a footnote to the end of the millennium's "theory of everything".

Jordan in his search for the basic principles which underlie QFT did not stay
with a quantization of oscillators setting as in the Dreim\"{a}nnerarbeit. In
his Habilitationsschrift \cite{Habil} 1927 he left no doubt that the crucial
point was not the quantization of a collection of oscillators, rather QFT was
the promotion of the causal locality principle of Maxwell's action at the
neighborhood which defines propagating classical field theory, to the realm of
quantum physics.

The distinction between the localization in QM based on Born's probability
applied to Schr\"{o}dinger wave functions, and on the other hand the causal
localization which is intrinsic to QFT without a directly associated
probability, is more radical than it appears at first sight. It accounts for
the significant difference between fields and particles in the presence of
interactions which remain unrelated within finite spacetime regions, but
become interwoven in large timelike regions through scattering theory. With
other words one needs the S-matrix to relate fields with particles, a relation
which for its mathematical formulation requires to study suitably defined
asmptotic limits of quantum fields. But even in the absence of interactions,
when particles and fields are directly related even in finite spacetime
regions, it would not be sufficient to tell a newcomer who just learned QM and
knows how to deal with arrays of uncoupled oscillators, that a "free field is
nothing but a collection of infinitely many harmonic oscillators"; he would
not know how to combine the infinite array of oscillators to a holistic object
which is localized in spacetime. To understand this holistic aspect took many
years from the inception of QFT in the Dreim\"{a}nnerarbeit to the
Jordan-Pauli work on causal commutator functions. And in the presence of
interactions the analogy with oscillators has anyhow no conceptual content.

Initially one may have thought that the distinction between QM and QFT is just
the existence of a maximal velocity in QFT (important for the notion of
micro-causality) compared to its absence in QM. As the example of acoustics
shows, special models of QM have an asymptotic (large timelike distances)
\textit{effective} limiting velocity obtained as a quantum mechanical
expectation value in appropriate states and depending on the kind of quantum
mechanical matter (viz. the acoustic velocity in QM). It is precisely in this
effective way that relativistic QM produces the velocity of light as an
effective limiting velocity, which is sufficient to construct a Poincar\'{e}
an invariant M\o ller operator and an S-matrix. Actually QFT employs both
concepts, for the asymptotic particle wave functions the frame-dependent
quantum mechanical localization related to the non-covariant position operator
and the causal micro-locality to be used for the description of local
observables and covariant fields. The latter played the crucial role in the
derivation of the experimentally verified dispersion relations\footnote{The
adaptation of the optical Kramers-Kronig relations to the scattering theory of
QFT was a very sophisticated rigorous theoretical project of the late 50s and
early sixties; in fact it was the only research topic in particle physics
which (together with is experimental verification) merits the characterization
"mission accomplished".}. The appearance of particles in the asymptotic
scattering limit of interacting theories turns out to lead to the
\textit{asymptotic coalescence of the probabilistic Born-Newton-Wigner
localization with the modular localization of QFT\footnote{There is no fight
of "Reeh-Schlieder (modular localization) wins against Newton-Wigner"
\cite{Hal} in large time asymptotic regions, rather only peaceful coexistence.
In particular the asymptotic movement of particles can be described by
velocity lines particle velocities being limited by the lightcone.}
\cite{interface}} in which the modular localization inherits the probability
and the the quantum mechanical BNW becomes covariant (independence of the
frame). This is also the only place where one needs to maintain a connection
between fields and particles. Without it QFT would just be an interesting
mathematical structure with no observational consequences.

One can safely assume that Jordan, at the time of his 1927 Habilitation, knew
some aspects of these propagation differences and their foundational
consequences. The first formula which incorporates the causal localization
structure of relativistic theories (as compares to the noncovariant equal time
commutation relations which also hold in nonrelativistic theories) appears in
his work with Pauli \cite{J-P}. The modern point of view that the different
models of QFTs are only different manifestations of a universal causal
locality principle of which the QFTs which can be be obtained by Lagrangian
quantization form only a small subset is an extension of that of Jordan but
one which is separated by many decades of hard work and a sociologically
caused loss of continuity with the first phase of QFT.

\section{Fields versus particles: Jordan and Dirac}

The issue of \textit{particles versus fields} is at the heart of relativistic
quantum physics and although amazing insights into this subtle issue have been
obtained since Jordan, nobody with a profound knowledge of modern QFT would
venture to say that it arrived at its conceptual closure. Without
interactions, i.e. for free fields, the two concepts are closely related, but
the presence of interaction-caused vacuum polarizations as a consequence of
relativistic localization separates them significantly. Although n-particle
states in the form of (anti)symmetrized tensor product of one-particle states
under certain special (but important) circumstances span the Hilbert space and
in this way grant the Hilbert space the form of a Wigner-Fock space, this
particle structure is basically global. On the other hand if in a model of QFT
for with a compact localized observable (i.e. a quantum field smeared with a
finite supported test function) applied to the vacuum leads to a to a
one-particle state, it is necessarily an interaction-free theory; or to phrase
it in the opposite way, the presence of an interaction is inexorably
accompanied by the presence of a vacuum polarization cloud (infinitely many
particle/antiparticle pairs) near the boundary of the spacetime localization
region inside which the desire one-particle state is localized. A similar
shorter but more drastic way to say the same is: a local "bang" on the vacuum
generates a vacuum polarization cloud.

The phenomenon of vacuum polarization was first seen with relativistic current
operators by Heisenberg \cite{Hei} and than generalized to interacting fields
in the setting of perturbation theory by Furry and Oppenheimer \cite{Fu-Op};
in the Dreim\"{a}nnerpapier QFT was basically the quantization rules of QT
applied to a chain of oscillators. It can be assumed that by 1930 Jordan had
at least some intuitive knowledge of the importance of vacuum polarizations in
connection with causal locality since he followed the development up to the
middle of the thirties when the worry about his less than satisfactory
position at a provincial university far from the ongoing quantum dialogue
which he enjoyed in Goettingen and the frustration about his inability to draw
some professional advantages from his support of the Nazi party begun to take
his better part \cite{Mainz}.

Before going into more details about the causal localization underlying QFT,
it is appropriate to present the easier part: the Born-Newton-Wigner
localization underlying QM and being the one which plays an important role in
assigning Born probabilities to wave functions of particles which is even
indispensable in QFT if it comes to the calculation of scattering cross
section. Whereas for finite times there is a significant distinction between
the frame-dependent Newton-Wigner localization (the Born-localization adapted
to the relativistic form of the inner product) and the covariant causal
localization defined through fields, their asymptotic large time compatibility
is the precondition for QFT in order to be an observational accessible.

\subsection{The position operator of QM, the frame-dependent particle
localization and the covariance of asymptotic correlations}

It is well-known that the localized states of Schr\"{o}dinger QM are
(improper) eigenstates of the position operator. They are frame-independent in
the Galileian sense of nonrelativistic QM. One can formally adapt this to
relativistic particle states (Newton-Wigner), but one looses the
frame-independence and hence the existence of a covariant position operator.
This localization appears in QFT only in an asymptotic sense, always in
connection with particles.

The \textit{direct particle interactions} (DPI) (where "direct" means "not
field-mediated") is a relativistic theory in the sense of representation
theory of the Poincar\'{e} group which, among other things, leads to a
Poincar\'{e} invariant S-matrix. Every property which can be formulated in
terms of particles (as the cluster factorization into systems with a lesser
number of particles as well as other timelike aspects of macrocausality), can
also be implemented in this setting. The S-matrix does however not fulfill
such analyticity properties as the crossing \cite{foun} property whose
derivation relies on the existence of local interpolating fields.

In contradistinction to the more fundamental locally covariant QFT, DPI is
primarily a phenomenological setting, but one which is consistent with every
property which can be expressed in terms of relativistic particles only. So
instead of approximating nonperturbative QFT outside conceptional-mathematical
control, the idea of DPI is to arrange phenomenological calculations in such a
way that the "approximation" at least preserves the principles of relativistic
mechanics and macro-causality i.e. of the only form of causality which can be
formulated in terms of interacting particles without invoking fields which
interpolate their incoming/outgoing asymptotic particles \cite{C-P}. Although
its practical use is limited to phenomenological settings for medium energy
nuclear processes in which only a few mesons are created, it is ideally suited
to understand what Dirac's relativistic particle view leads to (probably not
what he had in mind) and it also helps greatly why macrocausality in terms of
particles with an additional esthetical formal simplification led
St\"{u}ckelberg to Feynman rules \ before Feynman and before their field
theoretic derivation. This shows that formally there is a close relation of
the two settings, even though they are very far apart in their underlying
physical principles.

As for any quantum mechanical prescription, the necessity to choose
interaction functions (potentials V) instead of local coupling constants
between fields as in QFT, already indicates that such a description is less on
the fundamental and more on the phenomenological side. The characterization as
"phenomenological" should however not be confused with "ad hoc rules", which
often turns out to be the present day meaning of the word "effective"
interaction. In fact the methods of DPI particle interactions leading to a
unitary cluster-factorizing representation and a Poincar\'{e}-invariant
scattering matrix S are highly sophisticated since relativity contradicts the
naive implementation of the cluster factorization by simply adding pair
interactions of the two-particle subsystems. Perhaps if Dirac would have known
these intricacies he would not have insisted in a particle based approach up
to 1950 (when he finally took up QFT not only for photons but also for massive
quantum matter). But then we would have missed all the wonderful discoveries
he made under the pretense of relativistic particles which are not affected by
these subtleties of multiparticle interactions.

Needless to add the DPI and QFT are the only conceptual frameworks of
relativistic QT; DPI is a particle-based relativistic QM\footnote{By coupling
creation channels "by hand" one can extend the DPI setting to
creation/annihilation processes which still obey relativity and clustering;
but thetr is no vacuum polarization and the creation through scattering has to
be put in by hand instead of being a consequence of the localization principle
of QFT.} which besides the relativistically invariant semi-global M\o ller
operator and ensuing S-matrix contains no covariant objects at finite times
(in particular no covariant position operators), and covariant localizable QFT
which for implementing causal localization (related to the macroscopic maximal
velocity of light) requires the prize of containing no localized interacting
particles and admitting particle correlation probabilities only asymptotically.

For the interaction of two relativistic particles the introduction of
interactions amounted to add to the free mass operator (the Hamiltonian in the
c.m. system) an interact which depends on the relative position and momentum.
The exigencies of representation theory of the Poincar\'{e} group are then
fulfilled and the cluster property stating that $S\rightarrow\mathbf{1}$ for
large spatial separation is a consequence of the short ranged interaction.
Assuming for simplicity identical scalar Bosons, the c.m. invariant energy
operator is $2\sqrt{p^{2}+m^{2}}~$and the interaction is introduced by adding
an interaction term $v$%

\begin{equation}
M=2\sqrt{\vec{p}^{2}+m^{2}}+v,~~H=\sqrt{\vec{P}^{2}+M^{2}}%
\end{equation}
where the invariant potential $v$ depends on the relative c.m. variables $p,q$
in an invariant manner i.e. such that $M$ commutes with the Poincar\'{e}
generators of the 2-particle system which is a tensor product of two
one-particle systems.

One may follow Bakamjian and Thomas (BT) \cite{BT} and choose the Poincar\'{e}
generators in their way so that the interaction only appears in the
Hamiltonian. Denoting the interaction-free generators by a subscript $0,$ one
arrives at the following system of two-particle generators%
\begin{align}
\vec{K}  &  =\frac{1}{2}(\vec{X}_{0}H+H\vec{X}_{0})-\vec{J}\times\vec{P}%
_{0}(M+H)^{-1}\\
\vec{J}  &  =\vec{J}_{0}-\vec{X}_{0}\times\vec{P}_{0}\nonumber
\end{align}

The interaction $v$ may be taken as a \textit{local} function in the relative
coordinate which is conjugate to the relative momentum $p$ in the c.m. system;
but since the scheme anyhow does not lead to local differential equations,
there is not much to be gained from such a choice. The Wigner canonical spin
$\vec{J}_{0}$ commutes with $\vec{P}=\vec{P}_{0}$ and $\vec{X}=\vec{X}_{.0}$
and is related to the Pauli-Lubanski vector $W_{\mu}=\varepsilon_{\mu\nu
\kappa\lambda}P^{\nu}M^{\kappa\lambda}$ .

As in the nonrelativistic setting, short ranged interactions $v$ lead to
M\o ller operators and S-matrices via a converging sequence of unitaries
formed from the free and interacting Hamiltonian%
\begin{align}
\Omega_{\pm}(H,H_{0})  &  =\lim_{t\rightarrow\pm\infty}e^{iHt}e^{-H_{0}t}\\
\Omega_{\pm}(M,M_{0})  &  =\Omega_{\pm}(H,H_{0})\label{sec}\\
S  &  =\Omega_{+}^{\ast}\Omega_{-}\nonumber
\end{align}
The identity in the second line is the consequence of a theorem which says
that the limit is not affected if instead of $M$ one takes take a positive
function of $M$ (\ref{sec}) as $H(M),$ as long as $H_{0}$ is the same function
of $M_{0}.$ This insures the the asymptotic \textit{frame-independence of
objects as the M\o ller operators and the S-matrix }but not that of semi
asymptotic operators as formfactors of local operators between ket in and bra
out particle states. Apart from this \textit{identity for operators and their
positive functions} (\ref{sec}) which is not needed in the nonrelativistic
scattering, the rest behaves just as in nonrelativistic scattering theory. As
in standard QM, the 2-particle cluster property which says that $\Omega_{\pm
}^{(2)}\rightarrow\mathbf{1,}$ $S^{(2)}\rightarrow\mathbf{1,}$ if the two
particles which interact with short range interactions are increasingly
spacelike separated in the sense of the the centers of the wave packets.

The implementation of clustering is more delicate for three particles as can
be seen from the fact that the first attempts were started in 1965 by Coester
\cite{Coe} and considerably later generalized (in collaboration with Polyzou
\cite{C-P}) to an arbitrary high particle number. To anticipate the result
derived below, DPI leads to a consistent scheme which fulfills cluster
factorization but it has no useful second quantized formulation so it may
stand accused of lack of elegance; one is inclined to view less elegant
theories also as less fundamental. It is also more nonlocal and nonlinear than
QM, This had to be expected since adding interacting particles does not mean
adding up interactions as in Schr\"{o}dinger QM.

The BT form for the generators can be achieved inductively for an arbitrary
number of particles. As will be seen, the advantage of this form is that in
passing from n-1 to n-particles the interactions add after appropriate
Poincar\'{e} transformations to the joint c.m. system and in this way one ends
up with Poincar\'{e} group generators for an interacting n-particle system.
But for $n>2$ the aforementioned subtle problem with the cluster property
arises; whereas this iterative construction in the nonrelativistic setting
complies with cluster separability, this is not the case in the relativistic context.

This problem shows up for the first time in the presence of 3 particles
\cite{Coe}. The BT iteration from 2 to 3 particles gives the 3-particle mass operator%

\begin{align}
M  &  =M_{0}+V_{12}+V_{13}+V_{23}+V_{123}\label{add}\\
V_{12}  &  =M(12,3)-M_{0}(12;3),~M(12,3)=\sqrt{\vec{p}_{12,3}^{2}+M_{12}^{2}%
}+\sqrt{\vec{p}_{12,3}^{2}+m^{2}}\nonumber
\end{align}
and the $M(ij,k)$ result from cyclic permutations. Here $M(12,3)$ denotes the
3-particle invariant mass in case the third particle is a \textquotedblleft
spectator\textquotedblright,\ which by definition does not interact with 1 and
2. The momentum in the last line is the relative momentum between the
$(12)$-cluster and particle $3$ in the joint c.m. system and $M_{12}$ is the
associated two-particle mass i.e. the invariant energy in the $(12)$ c.m
system. Written in terms of the original two-particle interaction $v,$ the
3-particle mass term appears very nonlinear.

As in the nonrelativistic case, one can always add a totally connected
contribution. Setting this contribution to zero, the 3-particle mass operator
only depends on the two-particle interaction $v.~$But contrary to the
nonrelativistic case, the BT generators constructed with $M$ as it stands does
not fulfill the cluster separability requirement. The latter demands that if
the interaction between two clusters is removed, the unitary representation
factorizes into that of the product of the two clustersn i.e. one expects that
shifting the third particle to infinity will render it a spectator and result
in a factorization $U_{12,3}\rightarrow U_{12}\otimes U_{3}$. Unfortunately
what really happens is that the $(12)$ interaction also gets switched off in
this process i.e. $U_{123}\rightarrow U_{1}\otimes U_{2}\otimes U_{3}$ . The
reason for this violation of the cluster separability property, as a simple
calculation (using the transformation formula from c.m. variables to the
original $p_{i}$, i = 1, 2, 3) shows \cite{C-P}, is that, although the spatial
translation in the original system (instead of the $12,3$ c.m. system) does
remove the third particle to infinity as it should, unfortunately it also
drives the two-particle mass operator (with which it does not commute) towards
its free value which violates clustering.

In other words the BT produces a Poincar\'{e} covariant 3-particle interaction
which is additive in the respective c.m. interaction terms (\ref{add}), but
the Poincar\'{e} representation $U$ of the resulting system will not be
cluster-separable. However this is the time for intervention of a saving
grace: \textit{scattering equivalence}.

As shown first in \cite{Coe}, even though the 3-particle representation of the
Poincar\'{e} group arrived at by the above arguments violates clustering, the
3-particle S-matrix computed in the additive BT scheme turns out to have the
cluster factorization property. But without implementing the correct cluster
factorization also for the 3-particle Poincar\'{e} generators there is no
chance to proceed to a clustering 4-particle S-matrix.

Fortunately there always exist unitaries which transform BT systems into
cluster-separable systems \textit{without affecting the S-matrix}. Such
transformations are called \textit{scattering equivalences. }They were first
introduced into QM by Sokolov \cite{So} and their intuitive content is related
to a certain insensitivity of the scattering operator under quasilocal changes
of the quantum mechanical description at finite times. This is reminiscent of
the insensitivity of the S-matrix against local changes in the interpolating
field-coordinatizations\footnote{In field theoretic terminology this means
changing the pointlike field by passing to another (composite) field in the
same equivalence class (Borchers class) or in the setting of AQFT by picking
another operator from a local operator algebra.} in QFT by e.g. using
composites instead of the Lagrangian field.

The notion of scattering equivalences is conveniently described in terms of a
subalgebra of \textit{asymptotically constant operators} $C$ defined by%
\begin{align}
\lim_{t\rightarrow\pm\infty}C^{\#}e^{iH_{0}t}\psi &  =0\\
\lim_{t\rightarrow\pm\infty}\left(  V^{\#}-1\right)  e^{iH_{0}t}\psi &
=0\nonumber
\end{align}
where $C^{\#}$ stands for both $C$ and $C^{\ast}$. These operators, which
vanish on dissipating free wave packets in configuration space, form a
*-subalgebra which extends naturally to a $C^{\ast}$-algebra $\mathcal{C}$. A
scattering equivalence is a unitary member $V\in$ $\mathcal{C}$ which is
asymptotically equal to the identity (the content of the second line).
Applying this asymptotic equivalence relation to the M\o ller operator one obtains%

\begin{equation}
\Omega_{\pm}(VHV^{\ast},VH_{0}V^{\ast})=V\Omega_{\pm}(H,H_{0})
\end{equation}
so that the $V$ cancels out in the S-matrix. Scattering equivalences do
however change the interacting representations of the Poincar\'{e} group
according to $U(\Lambda,a)\rightarrow VU(\Lambda,a)V^{\ast}.$

The upshot is that there exists a clustering Hamiltonian $H_{clu}$ which is
unitarily related to the BT Hamiltonian $H_{BT}$ i.e. $H_{clu}=BH_{BT}B^{\ast
}$ such that $B\in\mathcal{C}~$is determined in terms of the scattering data
computed from $H_{BT}.$ It is precisely this clustering of $H_{clu}$ which is
needed for obtaining a clustering 4-particle S-matrix which is
cluster-associated with the $S^{(3)}$. With the help of $M_{clu}$ one defines
a 4-particle interaction following the additive BT prescription; the
subsequent scattering formalism leads to a clustering 4-particle S-matrix and
again one would not be able to go to n=5 without passing from the BT to the
cluster-factorizing 4-particle Poincar\'{e} group representation. Coester and
Polyzou showed \cite{C-P} that this procedure can be iterated and doing this
one arrives at the following statement

\textbf{Statement}: \textit{The freedom of choosing scattering equivalences
can be used to convert the Bakamijan-Thomas presentation of multi-particle
Poincar\'{e} generators into a cluster-factorizing representation. In this way
a cluster-factorizing S-matrix }$S^{(n)}$\textit{ associated to a BT
representation }$H_{BT}$\textit{ (in which clustering mass operator }%
$M_{clu}^{(n-1)}$\textit{ was used) leads via the construction of }%
$M_{clu}^{(n)}$\textit{ to a S-matrix }$S^{(n+1)}$\textit{ which clusters in
terms of all the previously determined }$S^{(k)},k<n.$ \textit{The use of
scattering equivalences prevents the existence of a 2}$^{nd}$\textit{
quantized formalism. }

For a proof we refer to the original papers \cite{C-P}\cite{P}. In passing we
mention that the minimal extension, i.e. the one determined uniquely in terms
of the two-particle interaction $v$) from n to n+1 for $n>3,$ contains
\textit{connected 3-and higher particle interactions} which are nonlinear
expressions involving nested roots in terms of the original two-particle
$v.~$This is another unexpected phenomenon as compared to the nonrelativistic case.

This theorem shows that it is possible to construct a relativistic theory
which however only uses particle concepts, thus correcting an old folklore
which says relativity + clustering = QFT. Whether one should call this DPI
theory "relativistic QM" or just a relativistic S-matrix theory in a QM
setting is a matter of taste; it depends on what significance one attributes
to those unusual scattering equivalences. In any case it defines a
\textit{relativistic S-matrix setting }with the correct particle behavior
i\textit{.}e\textit{. }all properties which one is able to formulate in terms
of particles (without the use of fields) as unitarity, Poincar\'{e} invariance
and macrocausality are fulfilled. In this context one should also mention that
the S-matrix bootstrap approach never addressed these macro-causality problems
of the DPI approach; it was a grand self-deluding design for a unique theory
of all non-gravitational interactions in which important physical details were
arrogantly ignored.

As mentioned above Coester and Polyzou also showed that this relativistic
setting can be extended to processes which maintain cluster factorization in
the presence of a finite number of creation/annihilation channels, thus
demonstrating, as mentioned before, that \textit{the mere presence of particle
creation is not characteristic for QFT} (but rather the presence of infinite
vacuum polarization clouds from "banging" with localized operators onto the
vacuum, see section 7). Different from the nonrelativistic Schr\"{o}dinger QM,
the superselection rule for masses of particles which results from Galilei
invariance for nonrelativistic QM does not carry over to the relativistic
setting; in this respect DPI is less restrictive than its Galilei-invariant QM
counterpart where such creation processes are forbidden.

One may consider the DPI setting of Coester and Polyzou as that scheme which
results from implementing the mentioned particle properties within a
n-particle Wigner representation setting in the presence of interaction
\cite{C-P}. Apparently the work of these mathematical nuclear physicists has
not been noted by particle physicists since the authors have published most of
their results in nuclear physics journals. What makes it worthwhile to mention
this work is that even physicists of great renown as Steven Weinberg did not
believe that such a theory exists because otherwise they would not have
conjectured that the implementation of cluster factorization properties in a
relativistic setting leads to QFT \cite{Wein}.

Certain properties which are consequences of locality in QFT and can be
formulated but not derived in a particle setting as the TCP symmetry, the
spin-statistics connection and the existence of anti-particles, can be added
"by hand" to the DPI setting. Other properties which are on-shell relics of
locality which QFT imprints on the S-matrix and which require the notion of
analytic continuation in particle momenta (as e.g. the crossing property for
formfactors) cannot be implemented in the QM setting of DPI.

\subsection{QFT and modular localization}

After having his feathers ruffled by Einstein on the issue of whether photons
(Nadelstrahlung) are necessary for acquiring radiation equilibrium, Jordan's
hopes that Einstein would support his beautiful solution in terms of field
quantization in the Dreim\"{a}nnerarbeit, which now supported Einstein's
photonsdid not materialize. He must have been somewhat disappointed with the
lukewarm reception of his calculation, which after all heralded the beginnings
of QFT in the same fateful year as Heisenberg proposed his QM. Einstein's
reluctance to embrace field quantization had of course nothing specifically to
do with Jordan's work in the Dreim\"{a}nnerarbeit, but rather with his own
reservation to except any form of quantum theory as the final description of
the new phenomena which had the Cain's mark of probability. Even Jordan's
collaborators in the Dreim\"{a}nnerarbeit had second thoughts about field
quantization\footnote{One later argument was "why should something which was
already quantum be quantized a second time" is justified with respect to QM
(no need to do this), but is based on a misunderstanding in the QFT context.}
for other reasons; but a critical attitude towards new speculative ideas
against a young iconoclast was quite normal; that the speculative method of
research at the frontier of particle theory requires a strong critical balance
was quite accepted. When Schweber pictures Jordan as the unsung hero
\cite{Schweber}, he referes to the lack of recognition\ two decades later when
most of his contemporaries received Nobel prizes except Jordan who after
Einstein Pauli and Heisenberg was one of the scientific giants of the century
before he left quantum physics and became involved in the turmoil of the Third Reich.

Before we comment on Jordan's view of QFT and its underlying causal locality
principle, as well as his unsucessful search for an intrinsic description
which does not follow the classical quantization parallellism but starts with
quantum principles\footnote{Jordan, as other physicists with a strong
philosophical background, did not accept that a less fundamental theory
(classical physics) via quantization calls the shots for a more basic one.},
we will remind the reader of the actual insight into this problem. In this way
it is easier to to understand the distance between Jordan's world of QFT and
the modern setting of an autonomous QFT.

Previously it was mentioned in passing that the localization underlying QFT
can be freed from the contingencies of field coordinatizations. This is best
achieved by a physically as well mathematically impressive recently discovered
structure of QFT which is still little known. Its name, "modular theory" is of
mathematical origin and refers to a vast generalization of the (uni)modularity
encountered in the relation between left/right Haar measure in group
representation theory. In the middle 60s the mathematician Tomita presented
this theory as a significant addition to the theory of operator algebras and
in the following years this theory received essential improvements from
Takesaki and later also from Connes.

At the same time Haag, Hugenholtz and Winnink published some work on
statistical mechanics of open systems \cite{Haag}. When physicists and
mathematicians using operator algebra methods in their research met at a
conference in Baton Rouge (Luisiana, USA) in 1966, there was mutual surprise
about the similarity of concepts, followed by deep appreciation of the
perfection with which these independently motivated developments supported
each other \cite{Borch}. Physicists not only adapted mathematical concepts
about operator algebras, but mathematicians also took some of their
terminology from physicists as e.g. \textit{KMS states} which refer to Kubo,
Martin and Schwinger who introduced an analytic property of thermal Gibbs
states merely as a computational tool (in order to avoid computing traces),
whereas Haag, Hugenholtz and Winnink realized that this property (the "KMS
property") plays a foundational conceptual role; it is the only property which
survives in the thermodynamic limit when the Gibbs trace formulas loses its
meaning and needs to be replaced by the analytic KMS boundary condition. After
the vacuum state, the KMS states are the most important states in QFT.

At that conference the relation of quantum field theoretical
\textit{localization} to the modular theory of operator algebras was still not
known, although Hawking's work on thermal aspects of physics inside and
outside a black hole event horizon could have already suggested that at the
localization behind an event horizon generates a situation with thermal
manifestations of localization are intimately related with aspects of this new
modular theory. The general connection of causal localization or localization
behind event horizons with KMS states and modular operator theory was made a
decade after Baton Rouge and directly after Hawking's work; first in a more
abstract paper in which the modular objects for wedge-localized algebras were
determined\footnote{The abstract Bisognano-Wichmann situation can be given a
more physical appearance in the form of the Unruh Gedankenexperiment
\cite{Unruh} of a uniformely accelerated observer whose world line is a
timelike hyperbola in the wedge-localized Rindler spacetime.} \cite{Bi-Wi}
being followed by arguments in favor of KMS thermal properties in black hole
physics really originating from the application of modular theory to quantum
matter enclosed behind an event horizon (the Schwarzschild horizon) \cite{Sew}.

The theory becomes more accessible if one introduces it first in its more
limited spatial- instead of its full algebraic- context. Since as a
foundational structure of LQP it merits more attention than it hitherto
received in the particle physics literature, some of its methods and
achievements will be presented in the sequel.

It has been realized by Brunetti, Guido and Longo\footnote{In a more limited
context and with less mathematical rigor this was independently proposed in
\cite{Sch}.} \cite{BGL} that there exists a natural intrinsic localization
structure on the Wigner representation space for any \textit{positive energy
representation} of the proper Poincar\'{e} group which is in particular
independent of the Born-Newton Wigner external localization structure (which
has no relation to the particular Wigner representation). The starting point
is an irreducible representation $U_{1}$of the Poincar\'{e}%
\'{}%
group on a Hilbert space $H_{1}$ that after "second quantization" becomes the
single-particle subspace of the Wigner-Fock Hilbert space $H_{WF}$ on which
the quantum fields act\footnote{The construction works for arbitrary positive
energy representations, not only for irreducible ones.}. In the bosonic case
the construction then proceeds according to the following steps \cite{BGL}%
\cite{Fa-Sc}\cite{MSY}.

One first fixes a reference wedge region, e.g. $W_{0}=\{x\in\mathbb{R}%
^{d},x^{d-1}>\left\vert x^{0}\right\vert \}$ and considers the one-parametric
L-boost group (the hyperbolic rotation by $\chi$ in the $x^{d-1}-x^{0}$ plane)
which leaves $W_{0}$ invariant; one also needs the reflection $j_{W_{0}}$
across the edge of the wedge (i.e. along the coordinates $x^{d-1}-x^{0}$). The
$\mathfrak{j}_{W_{0}}$-extended Wigner representation is then used to define
two commuting wedge-affiliated operators%
\begin{equation}
\mathfrak{\delta}_{W_{0}}^{it}=\mathfrak{u}(0,\Lambda_{W_{0}}(\chi=-2\pi
t)),~\mathfrak{j}_{W_{0}}=\mathfrak{u}(0,j_{W_{0}})
\end{equation}
where attention should be paid to the fact that in a positive energy
representation any operator which inverts time is necessarily
antilinear\footnote{The wedge reflection $\mathfrak{j}_{W_{0}}$ differs from
the TCP operator only by a $\pi$-rotation around the W$_{0}$ axis.}. A unitary
one- parametric strongly continuous subgroup as $\delta_{W_{0}}^{it}$ can be
written in terms of a selfadjoint generator $K$ as $\delta_{W_{0}}%
^{it}=e^{-itK_{W_{0}}}$ and therefore permits an "analytic continuation" in
$t$ to an unbounded densely defined positive operators $\delta_{W_{0}}^{s}$.
With the help of this operator one defines the unbounded antilinear operator
which has the same dense domain as its "radial" part%
\begin{equation}
\mathfrak{s}_{W_{0}}=\mathfrak{j}_{W_{0}}\mathfrak{\delta}_{W_{0}}^{\frac
{1}{2}},\text{ }\mathfrak{j\delta}^{\frac{1}{2}}\mathfrak{j}\mathfrak{=\delta
}^{-\frac{1}{2}} \label{c.r.}%
\end{equation}

Whereas the unitary operator $\delta_{W_{0}}^{it}$ commutes with the
reflection, the antiunitarity of the reflection changes the sign in the
analytic continuation which leads the commutation relation between $\delta$
and $\mathfrak{j}$ in (\ref{c.r.}). This causes the involutivity of the
s-operator on its domain, as well as the identity of its range with its
domain
\begin{align*}
\mathfrak{s}_{W_{0}}^{2}  &  \subset\mathbf{1}\\
dom~\mathfrak{s}  &  =ran~\mathfrak{s}%
\end{align*}
Such operators which \textit{are unbounded and yet involutive} on their domain
are very unusual; according to my best knowledge they only appear in modular
theory and it is precisely these unusual properties which are capable to
encode geometric localization properties into domain properties of abstract
quantum operators, a fantastic achievement completely unknown in QM. The more
general algebraic context in which Tomita discovered modular theory will be
mentioned later.

The involutivity means that the s-operator has $\pm1$ eigenspaces; since it is
antilinear, the +space multiplied with $i$ changes the sign and becomes the -
space; hence it suffices to introduce a notation for just one eigenspace%
\begin{align}
\mathfrak{K}(W_{0})  &  =\{domain~of~\Delta_{W_{0}}^{\frac{1}{2}%
},~\mathfrak{s}_{W_{0}}\psi=\psi\}\\
\mathfrak{j}_{W_{0}}\mathfrak{K}(W_{0})  &  =\mathfrak{K}(W_{0}^{\prime
})=\mathfrak{K}(W_{0})^{\prime},\text{ }duality\nonumber\\
\overline{\mathfrak{K}(W_{0})+i\mathfrak{K}(W_{0})}  &  =H_{1},\text{
}\mathfrak{K}(W_{0})\cap i\mathfrak{K}(W_{0})=0\nonumber
\end{align}

It is important to be aware that, unlike QM, we are here dealing with real
(closed) subspaces $\mathfrak{K}$ of the complex one-particle Wigner
representation space $H_{1}$. An alternative which avoids the use of real
subspaces is to directly work with complex dense subspaces as in the third
line. Introducing the graph norm of the dense space the complex subspace in
the third line becomes a Hilbert space in its own right. The second and third
line require some explanation. The upper dash on regions denotes the causal
disjoint (which is the opposite wedge) whereas the dash on real subspaces
means the symplectic complement with respect to the symplectic form
$Im(\cdot,\cdot)$ on $H_{1}.$

The two properties in the third line are the defining property of what is
called the \textit{standardness property} of a real
subspace\footnote{According to the Reeh-Schlieder theorem \cite{Haag} a local
algebra $\mathcal{A(O})$ in QFT is in standard position with respect to the
vacuum i.e. it acts on the vacuum in a cyclic and separating manner. The
spatial standardness, which follows directly from Wigner representation
theory, is just the one-particle projection of the Reeh-Schlieder property.};
any standard K space permits to define an abstract s-operator%
\begin{align}
\mathfrak{s}(\psi+i\varphi)  &  =\psi-i\varphi\\
\mathfrak{s}  &  =\mathfrak{j}\delta^{\frac{1}{2}}\nonumber
\end{align}
whose polar decomposition (written in the second line) yields two modular
objects, a unitary modular group $\delta^{it}$ and a antiunitary reflection
which generally have however no direct geometric significance. The domain of
the Tomita $\mathfrak{s}$-operator is the same as the domain of $\delta
^{\frac{1}{2}}$ namely the real sum of the K space and its imaginary multiple.
Note that this domain is determined solely in terms of Wigner group
representation theory.

It is easy to obtain a net of K-spaces by $U(a,\Lambda)$-transforming the
K-space for the distinguished $W_{0}.$ A bit more tricky is the construction
of sharper localized subspaces via intersections
\begin{equation}
\mathfrak{K}(\mathcal{O})=%
{\displaystyle\bigcap\limits_{W\supset\mathcal{O}}}
\mathfrak{K}(W)
\end{equation}
where $\mathcal{O}$ denotes a causally complete smaller region (noncompact
spacelike cone, compact double cone). Intersection may not be standard, in
fact they may be zero in which case the theory allows localization in $W$ (it
always does) but not in $\mathcal{O}.$ Such a theory is still causal but not
local in the sense that its generating free fields are pointlike. One can show
that the noncompact intersection for spacelike cones $\mathcal{O=C}$ for all
positive energy is always standard.

Note that the relativistic DPI setting also starts from Wigner particles, but
it ignores the presence of this autonomous and covariant modular localization
structure and instead introduces quantum mechanical frame-dependent
Born-Newton-Wigner localization based on position operators which have no
intrinsic relation to the Wigner representation theory of the Poincar\'{e}
group. It is easy to show that the dense subspaces $\mathfrak{K}%
\mathcal{(O)+}i\mathfrak{K}\mathcal{(O)}$ which are the domain of the
unbounded Tomita involution $\mathfrak{s}$ have all the properties demanded of
a net of $\mathcal{O}$-localized subspaces of the one-particle Wigner space.
The subtlety of causal localization here is, that whereas BNW localized
subspaces are (as spectral subspaces of a selfadjoint position operator)
closed subspaces, the frame-independent modular subspaces are dense and change
with the localization region; in fact the spacetime localization is fully
encoded in the dense domain of the modular $\mathfrak{s}(\mathcal{O})$ operator.

There are three classes of irreducible positive energy representation, the
family of massive representations $(m>0,s)$ with half-integer spin $s$ and the
family of massless representation which consists really of two subfamilies
with quite different properties, namely the $(0,h=~$half-integer$)$ class,
often called the neutrino-photon class, and the rather large class of
$(0,\kappa>0)$ infinite helicity representations parametrized by a
continuous-valued Casimir invariant $\kappa$ \cite{MSY}$.$

For the first two classes the $\mathfrak{K}$-space the standardness property
also holds for double cone intersections $\mathcal{O=D}$ for arbitrarily small
$\mathcal{D},$ but this is definitely not the case for the infinite helicity
family for which the localization spaces for compact spacetime regions turn
out to be trivial\footnote{It is quite easy to prove the standardness for
spacelike cone localization (leading to singular stringlike generating fields)
just from the positive energy property which is shared by all three families
\cite{BGL}.}. Passing from localized subspaces $\mathfrak{K}$ in the
representation theoretical setting to singular covariant generating wave
functions (the "first quantized" analogs of generating fields) one can show
that the $\mathcal{D}$ localization leads to pointlike singular generators
(state-valued distributions) whereas the spacelike cone localization
$\mathcal{C}$ is associated with semiinfinite spacelike stringlike singular
generators \cite{MSY}. Their "second quantized" counterparts are pointlike or
(in case of the infinite spin family) stringlike covariant fields. It is
remarkable that the modular localization concept does not require to introduce
generators which are localized on timelike hypersurfaces (branes).

Although the observation that the third Wigner representation class is not
pointlike generated was made many decades ago, the statement that it is
semiinfinite string-generated and that this is the worst possible case of
state localization is based on modular theory and is of a more recent vintage
\cite{BGL}\cite{MSY}.

There is a very subtle aspect of modular localization which one encounters in
the second Wigner representation class of massless finite helicity
representations (the photon, graviton class). Whereas in the massive case all
spinorial fields $\Psi^{(A,\dot{B})}$ the relation of the physical spin $s$
with the two spinorial indices follows the naive angular momentum composition
rules \cite{Wein}%
\begin{align}
\left\vert A-\dot{B}\right\vert  &  \leq s\leq\left\vert A+\dot{B}\right\vert
,\text{ }m>0\label{line}\\
s  &  =\left\vert A-\dot{B}\right\vert ,~m=0\nonumber
\end{align}
the second line contains the significantly reduced number of spinorial
descriptions for zero mass and finite helicity representations. What is going
on here, why is there, in contradistinction to classical field theory, no
covariant s=1 vector-potential $A_{\mu}$ or no $g_{\mu\nu}$ in case of s=2 ?
Why are the admissible covariant generators of the Wigner representation in
this case limited to field strengths (for s=2, the linearized Riemann tensor)
and why do their potentials inherit this localization only in the massive but
not in the massless case ?

The short answer is that there is really a deep clash between the quantum
aspects (Hilbert space structure) and spacetime localization which in the
($m=0,s\geq1$) family (different from the infinite spin family) does not
affect the representation as such but rather the existence of certain
generators with prescribed covariance properties (those not appearing in the
second line of (\ref{line})). In the gauge theory formulation one sacrifices
the Hilbert space and maintains (at least formally\footnote{From a physical
viewpoint the retention of the pointlike nature is a Pyrrhic victory since
this localization is void of physical meaning.}) a pointlike localization in
an indefinite metric Krein space whereas maintaining the Hilbert space means
relaxing the covariant localization from point to semiinfinite stringlike. But
this is not a bad compromise, the physical localization is really stringlike
generated. \ Taking this option, the full range of spinorial possibilities
(\ref{line}) returns in terms of string localized fields $\Psi^{(A,\dot{B}%
)}(x,e)$ if $s\neq\left\vert A-\dot{B}\right\vert $. These generating free
fields are covariant and "string-local"%

\begin{align}
U(\Lambda)\Psi^{(A,\dot{B})}(x,e)U^{\ast}(\Lambda)  &  =D^{(A,\dot{B}%
)}(\Lambda^{-1})\Psi^{(A,\dot{B})}(\Lambda x,\Lambda e)\label{string}\\
\left[  \Psi^{(A,\dot{B})}(x,e),\Psi^{(A^{\prime},\dot{B}^{\prime})}%
(x^{\prime},e^{\prime}\right]  _{\pm}  &  =0,~x+\mathbb{R}_{+}e><x^{\prime
}+\mathbb{R}_{+}e^{\prime}\nonumber
\end{align}
Here the unit vector $e$ is the spacelike direction of the semiinfinite string
and the last line expresses the spacelike fermionic/bosonic spacelike
commutation. The best known illustration is the ($m=0,s=1$) vectorpotential
representation; in this case it is well-known that although a generating
pointlike field strength exists, there is no \textit{pointlike}
vectorpotential acting in a Hilbert space.

According to (\ref{string}) the modular localization approach offers as a
substitute a stringlike covariant vector potential $A_{\mu}(x,e).$ In the case
($m=0,s=2$) the "field strength" is a fourth degree tensor which has the
symmetry properties of the Riemann tensor (it is often referred to as the
\textit{linearized} Riemann tensor). In this case the string-localized
potential is of the form $g_{\mu\nu}(x,e)$ i.e. resembles the metric tensor of
general relativity. Some consequences of this localization for a reformulation
of gauge theory will be mentioned in section 8.

Even in case of massive free theories where the representation theoretical
approach of Wigner does not require to go beyond pointlike localization,
covariant stringlike localized fields exist. Their attractive property is that
they improve the short distance behavior e.g. a massive pointlike
vector-potential of \textit{sdd=2} passes to a string localized vector
potential of \textit{sdd=1}. In this way the increase of the sdd of pointlike
fields with spin s can be traded against string localized fields of spin
independent dimension with sdd=1. This observation would suggest the
possibility of an enormous potential enlargement of perturbative accessible
higher spin interaction in the sense of power counting.

A different kind of spacelike string-localization arises in d=1+2 Wigner
representations with anomalous spin \cite{Mu1}. The amazing power of the
modular localization approach is that it preempts the spin-statistics
connection already in the one-particle setting, namely if s is the spin of the
particle (which in d=1+2 may take on any real value) then one finds for the
connection of the symplectic complement with the causal complement the
generalized duality relation
\begin{equation}
\mathfrak{K}(\mathcal{O}^{\prime})=Z\mathfrak{K}(\mathcal{O})^{\prime}%
\end{equation}
where the square of the twist operator $Z=e^{\pi is}~$is easily seen (by the
connection of Wigner representation theory with the two-point function) to
lead to the statistics phase $=Z^{2}$ \cite{Mu1}.

The fact that one never has to go beyond string localization (and fact, apart
from $s\geq1$, never beyond point localization) in order to obtain generating
fields for a QFT is remarkable in view of the many attempts to introduce
extended objects into QFT.

It is helpful to be again reminded that modular localization which goes with
real subspaces (or dense complex subspaces), unlike B-N-W localization, cannot
be connected with probabilities and projectors. It is rather related to causal
localization aspects; the standardness of the K-space for a compact region is
nothing else then the one-particle version of the Reeh-Schlieder property. As
will be seen in the next section modular localization is also an important
tool in the non-perturbative construction of interacting models.

A net of real subspaces $\mathfrak{K}(\mathcal{O})$ $\subset$ $H_{1}$ for an
finite spin (helicity) Wigner representation can be "second
quantized"\footnote{The terminology 2$^{nd}$ quantization is a misdemeanor
since one is dealing with a rigorously defined functor within QT which has
little in common with the artful use of that parallellism to classical theory
called "quantization". In Edward Nelson's words: (first) quantization is a
mystery, but second quantization is a functor.} via the CCR (Weyl)
respectively CAR quantization functor; in this way one obtains a covariant
$\mathcal{O}$-indexed net of von Neumann algebras $\mathcal{A(O)}$ acting on
the bosonic or fermionic Fock space $H=Fock(H_{1})$ built over the
one-particle Wigner space $H_{1}.$ For integer spin/helicity values the
modular localization in Wigner space implies the identification of the
symplectic complement with the geometric complement in the sense of
relativistic causality, i.e. $\mathfrak{K}(\mathcal{O})^{\prime}%
=\mathfrak{K}(\mathcal{O}^{\prime})$ (spatial Haag duality in $H_{1}$). The
Weyl functor takes this spatial version of Haag duality into its algebraic
counterpart. One proceeds as follows: for each Wigner wave function
$\varphi\in H_{1}$ the associated (unitary) Weyl operator is defined as%
\begin{align}
Weyl(\varphi)  &  :=expi\{a^{\ast}(\varphi)+a(\varphi)\}\in B(H)\\
\mathcal{A(O})  &  :=alg\{Weyl(\varphi)|\varphi\in\mathfrak{K}(\mathcal{O}%
)\}^{^{\prime\prime}},~~\mathcal{A(O})^{\prime}=\mathcal{A(O}^{\prime
})\nonumber
\end{align}
where $a^{\ast}(\varphi)$ and $a(\varphi)$ are the usual Fock space creation
and annihilation operators of a Wigner particle in the wave function $\varphi
$. We then define the von Neumann algebra corresponding to the localization
region $\mathcal{O}$ in terms of the operator algebra generated by the
functorial image of $\mathfrak{K}(\mathcal{O})$ as in the second line. By the
von Neumann double commutant theorem, our generated operator algebra is weakly closed.

The functorial relation between real subspaces and von Neumann algebras via
the Weyl functor preserves the causal localization structure and hence the
spatial duality passes to its algebraic counterpart. The functor also commutes
with the improvement of localization through intersections $\cap$ according to
$\mathfrak{K}(\mathcal{O})=\cap_{W\supset O}\mathfrak{K}(W),~\mathcal{A(O}%
)=\cap_{W\supset O}\mathcal{A}(W)$ as expressed in the commuting diagram%
\begin{align}
&  \left\{  \mathfrak{K}(W)\right\}  _{W}\longrightarrow\left\{
\mathcal{A}(W)\right\}  _{W}\\
&  \ \ \downarrow\cap~~~\ \ \ \ \ \ \ \ \ \ ~\ ~\downarrow\cap\nonumber\\
~~  &  \ \ \ \mathfrak{K}(\mathcal{O})\ \ \ \longrightarrow\ \ ~\mathcal{A(O}%
)\nonumber
\end{align}
Here the vertical arrows denote the tightening of localization by intersection
whereas the horizontal ones indicate the action of the Weyl functor.

The case of half-integer spin representations is analogous \cite{Fa-Sc}, apart
from the fact that there is a mismatch between the causal and symplectic
complements which must be taken care of by a \textit{twist operator}
$\mathcal{Z}$ and as a result one has to use the CAR functor instead of the
Weyl functor.

In case of the large family of irreducible zero mass infinite spin
representations in which the lightlike little group is faithfully represented,
the finitely localized K-spaces are trivial $\mathfrak{K}(\mathcal{O})=\{0\}$
and the \textit{most tightly localized nontrivial spaces} \textit{are of the
form} $\mathfrak{K}(\mathcal{C})$ for $\mathcal{C}$ an arbitrarily narrow
\textit{spacelike cone}. As a double cone contracts to its core which is a
point, the core of a spacelike cone is a \textit{covariant spacelike
semiinfinite string}. The above functorial construction works the same way for
the Wigner infinite spin representation, except that in that case there are no
nontrivial algebras which have a smaller localization than $\mathcal{A(C})$
and there is no field which is sharper localized than a semiinfinite string.
As stated before, stringlike generators, which are also available in the
pointlike case, turn out to have an improved short distance behavior which
makes them preferable from the point of view of formulating interactions
within the power counting limit. They can be constructed from the unique
Wigner representation by so called intertwiners between the unique canonical
and the many possible covariant (dotted-undotted spinorial) representations.
The Euler-Lagrange aspects plays no direct role in these construction since
the causal aspect of hyperbolic differential propagation are fully taken care
of by modular localization and also because most of the spinorial higher spin
representations (\ref{line}) cannot be characterized in terms of
Euler-Lagrange equations. Modular localization is the more general method of
implementating causal propagation than that in terms of hyperbolic equations
of motions.

A basis of local covariant field coordinatizations is then defined by the free
field and its Wick-ordered composites. The case which deviates furthest from
classical behavior is the pure stringlike infinite spin case which relates a
\textit{continuous} family of free fields with one irreducible infinite spin
representation. Its non-classical aspects, in particular the absence of a
Lagrangian, is the reason why the spacetime description in terms of
semiinfinite string fields has been discovered only recently rather than at
the time of Jordan's field quantization or Wigner's representation theoretical approach.

Using the standard notation $\Gamma$ for the second quantization functor which
maps real localized (one-particle) subspaces into localized von Neumann
algebras and extending this functor in a natural way to include the images of
the $\mathfrak{K}(\mathcal{O})$-associated $s,\delta,j$ which are denoted by
$S,\Delta,J,$ one arrives at the Tomita Takesaki theory of the
interaction-free local algebra ($\mathcal{A(O}),\Omega$) in standard
position\footnote{The functor $\Gamma$ preserves the standardness i.e. maps
the spatial one-particle standardness into its algebraic counterpart.}%
\begin{align}
&  H_{Fock}=\Gamma(H_{1})=e^{H_{1}},~\left(  e^{h},e^{k}\right)
=e^{(h,k)}\label{mod}\\
&  \Delta=\Gamma(\delta),~J=\Gamma(j),~S=\Gamma(s)\nonumber\\
&  SA\Omega=A^{\ast}\Omega,~A\in\mathcal{A}(O),~S=J\Delta^{\frac{1}{2}%
}\nonumber
\end{align}
This result is a special case of the Tomita-Takesaki theorem which is a
statement about the existence of two modular objects $\Delta^{it}$ and $J$ on
the algebra%
\begin{align}
\sigma_{t}(\mathcal{A(O}))  &  \equiv\Delta^{it}\mathcal{A(O})\Delta
^{-it}=\mathcal{A(O})\\
J\mathcal{A(O})J  &  =\mathcal{A(O})^{\prime}=\mathcal{A(O}^{\prime})\nonumber
\end{align}
in words: the reflection $J$ maps an algebra (in standard position) into its
von Neumann commutant and the unitary group $\Delta^{it}$ defines an
one-parametric automorphism-group $\sigma_{t}$ of the algebra. In this form
(but without the last geometric statement involving the geometrical causal
complement $\mathcal{O}^{\prime})$ the theorem hold in complete mathematical
generality for standard pairs ($\mathcal{A},\Omega$). The free fields and
their Wick composites are "coordinatizing" singular generators of this
$\mathcal{O}$-indexed net of operator algebras in that the smeared fields
$A(f)$ with $suppf\subset\mathcal{O}$ are (unbounded operators) affiliated
with $\mathcal{A(O})$ and in a certain sense generate $\mathcal{A(O}).$

In the above second quantization context the origin of the T-T theorem and its
proof is clear: the symplectic disjoint passes via the functorial operation to
the operator algebra commutant (\ref{line}) and the spatial one-particle
automorphism goes into its algebraic counterpart. The definition of the Tomita
involution $S$ through its action on the dense set of states (guarantied by
the standardness of $\mathcal{A}$) as $SA\Omega=A^{\ast}\Omega$ and the action
of the two modular objects $\Delta,J$ (\ref{mod}) is however part of the
general setting of the modular Tomita-Takesaki theory of abstract operator
algebras in "standard position"; standardness is the mathematical terminology
for the physicists Reeh-Schlieder property i.e. the existence\footnote{In QFT
any finite energy vector (which of course includes the vacuum) has this
property as well as any nondegenerated KMS state. In the mathematical setting
it is shown that standard vectors are "$\delta-$dense" in $H$.} of a vector
$\Omega\in H$ with respect to which the algebra acts cyclic and has no
"annihilators" of $\Omega.$ Naturally the proof of the abstract T-T theorem in
the general setting of operator algebras is more involved\footnote{The local
algebras of QFT are (as a consequence of the split property) hyperfinite; for
such operator algebras Longo has given an elegant proof \cite{hyper}.}. It
validity can be established in interacting QFT either in a Wightman
\cite{Wight} setting or in theories which have a complete scattering
interpretation \cite{Muscat}

The domain of the unbounded Tomita involution $S$ turns out to be
"kinematical" in the sense that the dense set which features in the
Reeh-Schlieder theorem is determined in terms of the representation of the
connected part of the Poincar\'{e} group i.e. the particle/spin
spectrum\footnote{For a wedge $W$ the domain of $S_{W}$ is determined in terms
of the domain of the "analytic continuation" $\Delta_{W}^{\frac{1}{2}}$ of the
wedge-associated Lorentz-boost subgroup $\Lambda_{W}(\chi),$ and for subwedge
localization regions $\mathcal{O}$ the dense domain is obtained in terms of
intersections of wedge domains.}. In other words the Reeh-Schlieder domains in
an interacting theory with asymptotic completeness are identical to those of
the incoming or outgoing free field theory.

The important property which renders this useful beyond free fields as a new
constructive tool in the presence of interactions, is that for $\left(
\mathcal{A}(W),\Omega\right)  ~$ the antiunitary involution $J$ depends on the
interaction, whereas $\Delta^{it}$ continues to be uniquely fixed by the
representation of the Poincar\'{e} group i.e. by the particle content. In fact
it has been known for some \cite{Sch} time that $J$ is related with its free
counterpart $J_{0}$ through the scattering matrix%
\begin{equation}
J=J_{0}S_{scat} \label{scat}%
\end{equation}

This modular role of the scattering matrix as a relative modular invariant
between an interacting theory and its free counterpart comes as a surprise. It
is precisely this property which opens the way for an inverse scattering
construction. Hence the properties of $J$ are essentially determined by the
relation of localized operators $A~$to their Hermitian adjoints $A^{\ast}%
$\footnote{According to a theorem of Alain Connes \cite{Connes} the existence
of operator algebras in standard position can be inferred if the real subspace
$K$ permit a decompositions into a natural positive cone and its opposite with
certain facial properties of positive subcones. Although this construction has
been highly useful in Connes classification of von Neumann factors, it has not
yet been possible to relate this to physical concepts.}.

Jordan had an intuitive understanding of causal localization in analogy to the
causal propagation (Cauchy propagation) in classical field theories as e.g.
the Maxwell theory. But his observation that in gauge theories the gauge
invariant physical matter fields were not point- but string- localized
(section 6) was not pursued to its foundational roots.

The string localization of potentials plays an essential role in understanding
the origin the nonlocality of charged and colored quantum matter as a result
of its interaction with nonlocal potentials. In this way the
Dirac-Jordan-Mandelstam "string" ( see section6 (\ref{DJM})) is replaced by
the matter field in a new perturbation theory in which "gauge" has been
replaced by noncompact stringlike localization in a Hilbert space (no
indefinite metric) setting such that the compact localized pointlike generated
subalgebra coincides with the gauge invariant subalgebra of the standard gauge
approach \cite{char}.

An important mechanism which has been almost forgotten in the present
discourse about the expectations from the impending LHC experiments is the
Schwinger-Higgs screening mechanism. The idea is that gauge theory with scalar
matter has a screened counterpart. This setting has conceptual differences
with the Higgs mechanism. There is a clear distinction to the Goldstone
spontaneous symmetry breaking; whereas for the latter the conserved current
leads to an infinite charge (large distance divergence from presence of
Goldstone Boson), the screening mechanism leads to a vanishing charge and
conversion of complex matter into real matter. The ensuing loss of the charge
superselection rule leads to a loss of symmetry which affects even the
even/odd symmetry of the remaining matter field \cite{char}.

\subsection{Speculative ideas without critical counter-balance}

QFT was born almost simultaneous with QM in a critical surrounding. Not only
did it have to assert itself against older proposals based on Bohr's model but
without photons, but even Einstein's support was muffled, and Jordan's
coauthors Born and Heisenberg maintained a critical detachment, at least for
some time. Dirac was one of the first who embraced QM and became the leading
figure in its shaping, but he needed almost 2 decades to wholeheartedly
embrace QFT as a unifying principle in relativistic particle theory. The first
pre-renormalization textbooks by Wenzel and Heitler used the QFT terminology,
but employed in large parts the setting of relativistic QM which the lowest
so-called tree approximation is nearly indistinguishable from QFT. The litmus
test for QFT versus relativistic QM came with the problem of how to treat
interaction-induced vacuum polarization i.e. how to cope with renormalization
theory (higher order charged loops) which is the heart piece of the
perturbative approach to QFT and the property which sharply separates QFT from
relativistic QM.

In the later part of his life Jordan's research was limited to classical
general relativity as well as pure mathematical problems of algebraic
structures. In this way the postwar development of renormalized perturbation
theory of QED as well as the subsequent derivation of dispersion relations in
particle theory (which vindicated the causality ideas of Jordan for the first
time in the laboratory) took place outside his attention. Even though Jordan
was the first to transfer the ideas of classical gauge theory to the realm of
quantum physics, he did not take notice of the renaissance of gauge theory
culminating in the standard model.

A less balanced train of thoughts started with S-matrix theory. The S-matrix
is, whenever it exists, a global (nonlocal) object which can be computed from
the fields, but in contrast to the letter it is free of vacuum polarization
and therefore also free of ultraviolet divergences. \ Another reason came into
the foreground after renormalized perturbation theory took some of the menace
out of the ultraviolet divergence issue. It was the failure of perturbative
QFT to account for the strong nuclear interactions; as a result the attention
shifted to more phenomenological methods closer to experimental accessible
quantities as scattering cross sections. If it comes to nuclear interactions
hardly anybody would claim that a nucleon or meson field is a measurable
quantity of direct physical significance.

The emphasis on the S-matrix forced particle theoreticians to think somewhat
harder about the field-particle relation with the result that, although
particle states are important states in the Hilbert space\footnote{In the most
important case of asymptotic completeness the Hilbert space even has the
structure of a Wigner-Fock space multiparticle space.}, which admit an
asymptotic relation to fields through the LSZ scattering theory, there is no
direct field-particle relation in finite localization regions. QFT demystified
the \textit{particle-wave duality} (e.g. the Schr\"{o}dinger \textit{wave}
functions describing \textit{states of particles}), but although the quite
different \textit{particle-field relation} is much better understood than at
the time of Jordan, it did not yet arrive at its closure and it may well turn
out that the complete understanding of this relation is identical to the
closure of QFT. But as a rule of thumb in interacting QFT: \textit{the larger
the spacetime localization region }$O$\textit{, the easier it is to control
the vacuum polarization admixtures} to particle states obtained from applying
operators $A\mathcal{(O)}$ to the vacuum $\Omega.$

The first application of these results to the mentioned particle-adapted
Kramers-Kronig dispersion theory was a resounding success. In fact it was a
project of considerable foundational value in that the causal locality concept
(on which QFT, in contradistinction to QM, is based on) was put to a test up
to the at that time highest available energies way beyond those obtained in
QED experiments one decade before. This project was well-defined and limited
in scope. In contradistinction to other high energy experiments this was not a
check on a particular model, but rather on a principle which was shared by all
models of QFT but not by QM. Without a rigorous mathematical-conceptual
derivation of these relations from the causality and spectral principles of
QFT, the subsequent experimental verification would not have had the enormous
significance for the continued credibility in the causal locality principles
of QFT lasting into the present. This research required an S-matrix setting
with the dispersion theoretic analytic properties (the relation between the
real and the absorptive part of the two-particle scattering amplitude) coming
from the causality principle of QFT. This should be distinguished from a pure
S-matrix approach.

The project attracted the attention of the leading physicists in the 50s and
60s and it was successfully completed after less than a decade; in fact it is
the only project ever in particle theory which ended with a "mission
accomplished" seal. All other projects are either ongoing at a very slow pace
(the standard model research) or entangled themselves in conceptual
contradiction as the post dispersion relation S-matrix setting starting with
the S-matrix bootstrap, passing to the dual model which in turn led into the
so-called string theory (no relation to the QFT strings of the previous
section!) which together with supersymmetry dominated the thinking of many
particle theorists. These developments never arrived at a theoretically
consistent setting nor did any important observational accessible prediction
arise from the almost 50 years dominance of these ideas.

In order to be not misunderstood, a foundational theory should not be
subjected to a time limit and neither should it be terminated for its lack of
observational success. After all, what is at stake according to the followers
of string theory, is a third kind of relativistic QT on the side of Jordan's
QFT and Dirac's relativistic QM (or DPI), a theory which is claimed to be
string-localized and rather unique and incorporates gravity (a "theory of
everything") and extends QFT. In the good times of particle theory these
claims would have been more credible because there would have been people like
Oppenheimer, Pauli, Feynman, Schwinger, Lehmann, Jost, K\"{a}ll\'{e}n...who
would have looked at them in a very critical manner. What really caused
considerable damage to particle physics is that the critical balance broke
down and only historians may find out whether this was caused by the
ideological zeal of the string community or the ivory tower mentality of those
who could have known better.

Even if nature would not make use of such a theory, the
conceptual-philosophical implication of its mere existence would be startling
and certainly reveal new aspects of QFT. Although QFT can be characterized as
the theory which realizes the principle of modular localization, there are
many aspects between the principles and the properties of models which we do
not know yet; contrary to the claim of some string theorists, their opinion
that QFT has reached its closure and is superseeded by string theory is an
illusion resulting from the belief that it is limited to what one finds about
it in textbooks.

Before we look at the nature and the causes for the conceptual derailments of
the S-matrix based dual model and of string theory in particular, it is
important to emphasize that this has nothing to do with a decrease of the
intellectual capacity as compared to authors working in QFT at the time of
Jordan, it is rather related to the loss of the delicate equilibrium between
speculative exploration at the frontiers of research and its critical
reflection. In a globalized community whose main purpose is to do research on
a scientific monoculture and strengthen its hegemony, there is no place for a
culture of autocriticism. Whereas an individual researcher would react to
critique with attention and sensitivity, somebody embedded in such a community
typically acts according to the vernacular that "that many cannot err". For
people who think that I am exaggerating, my recommendation is to read what a
critical insiders has to say about the situation \cite{Zap}.

In the following I will analyze two closely interconnected claims which
dominated relativistic QT for many decades. They address the central issue of
relativistic localization which from the time of Jordan to the present has
grown in relevance and is now \textit{the} defining property of QFT (previous
subsection). Both theories, namely string theory and its predecessor the dual
model, contain incurable conceptual errors resulting from a misunderstanding
of localization and the consequences of stringlike localization on the
S-matrix. \ Since the lessons to be learned from the exposition of error on a
subtle problem are often not less important as those from a valid theory, the
remainder of this subsection will contain detailed conceptual-mathematical
arguments why the following two claims which undermine the ideas around the
dual model and string theory are valid.

1) String theory as obtained by canonical quantization of the Nambu-Goto
Lagrangian does not deal with string-localized objects in spacetime but rather
a free string describes a "dynamically infinite component pointlike field"

2) A multicomponent chiral conformal theory has no string-like embedding into
its "target space" i.e. its inner symmetry index space even in those cases in
which the inner symmetry group of the chiral theory can be interpreted as a
space on which the Poincar\'{e} group acts.

The first wrong track concerning the localization of the quantum theory of the
canonical quantized bilinearized\footnote{The original Lagragian in terms of
the square root of the surface element is an integrable system unrelated to
string theory as we know it \cite{Pohl}. Its classical solutions can be
related to a linear equation of motion structure which in turn may be encoded
into a bilinear Lagrangian which is susceptible to canonical quantization.}
Nambu-Goto Lagrangian \cite{Na-Go} came from the analogy%
\begin{align}
\mathcal{L}_{part}  &  =-mc\sqrt{ds^{2}}\curvearrowright\text{ }%
relativistic\text{ }particle\\
\mathcal{L}_{string}  &  =-\frac{1}{2\pi\alpha}\sqrt{dArea}\curvearrowright
\text{ }relativistic~string\nonumber
\end{align}
An incorrect idea already entered into the first line. Although metaphoric
pictures are sometimes helpful, this one is patently wrong. There are no
covariant position operators and a fortiori no quantum mechanical covariant
world line; even in a so-called relativistic QM (DPI) it is only the M$o$ller
operator and the S-matrix which are invariant, but there are no Lorentz
covariant (spinorial) operators. The first who realized this was Jordan's
contemporary Eugen Wigner. In fact the correct description for relativistic
particles is in terms of Wigner's representation theoretical classification,
whereas the above metaphoric description was only invented as a confidence
inspiring metaphor by string theorist. Needless to add that the mathematical
problems related to the "reparametrization invariance" of such a square root
action create infinities whose removal remains part of metaphoric rather than
of mathematical physics. Subscribing to such a picture of particles does not
only mean abandoning Wigner's crystal clear particle classification theory,
but it also leads to a conflict with the notion of relativistic localization.
This is certainly not a helpful analogy to string theory.

It is perfectly possible to calculate the wave function space and the
associated free "string" without relying in such doubtful metaphors. This was
first done by string theorists \cite{Mar}\cite{Lowe} and the result was that
the (graded in the supersymmetric case) \textit{commutator is that of an
infinite component pointlike-localized field}. Unfortunately the community
pressure had already created a scissor in the head of these authors so that
they simply declared the point of localization as lying on a (presumably
invisible) string. Who would have expected an affirmative conclusion from a
correct calculation against the prevailing view of their string community was
left disappointed. A further study of the sociological mechanisms which
operate in a situation where even a correct calculation serves to support
incorrect ideas, should be left to future historians and philosophers of
science. Errors occur wherever humans work on innovative problems and can even
be found in Jordan's oeuvre, however nowadays the protective community aspects
adds a different dimension by globalizing them.

The incorrect localization assignment did not start with ST; in a slightly
different form it was already present in the dual model. This is obvious in
the operator representations of this model, which consists in expressing the
would be scattering amplitudes in terms of operators from a multi-component
chiral conformal theory, in which the chiral fields are formally exponential
of potentials $\Phi_{k}$ of an n-component abelian current model\footnote{The
exponential model has therefore the form of an n-component conformal nonlinear
sigma model on which the n-component Poincar\'{e} acts in the form of an inner
symmetry \cite{Vech}
\par
.}. The dual model reading interprets the anomalous scale dimensions of these
objects as masses of particles and the numerical vector $\vec{\alpha}$ in the
exponential $exp\sum i\vec{\alpha}\vec{\Phi}$ as a particle momentum whereas
the chiral field $\Phi_{k}(\tau)$ k=1...$n$ is assigned the role of a position
$X_{k}(\tau)$ operator in an n-dimensional space which traces out a string as
$\tau$ runs around a circle (taking the compact description of chiral
conformal QFT). With other words this object is thought of as defining an
embedding of a string into the n-dimensional component space called "the
target space". Apart from a misunderstanding of field theoretic localization
in an imagined setting of relativistic position operators, there is of course
also the extremely strange idea of embedding a QFT into what is normally
called its internal symmetry space.

To make a long story short, it is possible to find a representation of the
Poincar\'{e} group in the internal symmetry space of a chiral theory and if
one requires this representation to be unitary and of positive energy there is
essentially only one solution: the so called n=10 superstring (which requires
some of the $\vec{\Phi}$ components to be spinorial instead of vectorial). But
the result describes (as in the previous case of the Nambu-Goto Lagrangian) a
pointlike object; the stringlike nature was, as in the previous case, just a
metaphoric conjecture. In fact the localization concept of QFT simply does not
permit an embedding of a lower into a higher dimensional theory; the
aforementioned stringlike fields do not result from an imbedding of a one
dimensional chiral theory into a higher dimensional space.

In this respect localization in QFT is completely different from that in the
better known QM. Whereas in QM one can simply take a one-dimensional string
and promote it to a string in higher dimensions by adding coordinates, this is
not possible since the causal localization is a holistic concept which is
already evident from the fact that the restriction of the vacuum state to the
observables localized in a spacetime regions is a highly subtle change in that
on the smaller collection of observables in that spacetime region leads to the
vacuum loosing its purity property and becoming a singular (no density matrix
description) KMS thermal state\footnote{The KMS Hamiltonian is only physical
if the localization comes from an objective event horizons of black holes
rather than being an observer.dependent "Gedanken-localization" in Minkowski
spacetime.}. In fact its holistic nature is the reason why geometry as a
mathematical discipline and the holistic geometry based on modular quantum
localization remain two different pairs of shoes. The first is neutral with
respect to applications\footnote{For example the theory of Riemann surfaces
has applications in 3-dim. geometry, the theory of analytic functions, theory
of Fuchsian groups etc. Saying however that a chiral model of QFT lives on a
Riemann surface is misleading since its physical localization is on
one-dimensional submanifolds.} whereas the is QFT context-dependent. This also
explains why attempts to let QFT dance according to the tune of geometric
mathematical ideas does not really clarify physical aspects.

An interesting historical illustration of the weakness of a mere geometric
point of view is the Wess-Zumino-Witten-Novikov action. Before it acquired
this long name, these models had been dealt with representation theoretical
methods of current algebras in the setting of a "multicomponent Thirring
model". At this level an association with a Lagrangian and a euclidean action
was unclear and also not needed since a representation theoretical approach is
intrinsic and self-sufficient. The only thing which a Lagrangian may add is
compact elegant presentation by reading it back into classical field theory.
Although Witten gave a positive answer \cite{WZW} (and Novikov added
mathematical insight) to this question in the form of a conformal sigma-model
field, the occurrence of a topological 3-dimensional euclidean boundary term
whose influence on spacetime localization remained obscure and separates this
topological Lagrangian from standard Lagrangians which are susceptible to
perturbative approximations. Fact is that the only constructive results on
two-dimensional conformal \cite{Lo-Ka} or massive nontrivial models
\cite{Lech} did not come from classically inspired geometric Lagrangian or
functional ideas but rather from algebraic representation-theoretic methods
based on modular localization.

On the other hand there is no other formalism in QT which connects so seamless
to the old quasiclassical Bohr-Sommerfeld QT and perturbation theory as the
Lagrangian- and functional integral. setting; the transition to the full QM
would have been a home game in the setting of Feynman's pathe integral without
going through all unaccustomed new nongeometric concepts in the work of
Heisenberg and the Dreim\"{a}nnerarbeit. But perhaps we would not have been
that happy with such a course of events, because although to reproduce
quasiclassical results and arrive at some perturbative computations would have
been simple, the exact calculation of integrable systems as the hydrogen atom
would have remained inaccessible, not to speak about the important concepts as
operators, states and the Born probabilty which the path integral does not
deliver by itself and which only with sufficient hindsight can be extracted.

The situation becomes worse in QFT where such representations also have an
intuitive metaphoric appeal\footnote{One obtains the correct perturbative
combinatoric and with a bit of hindsight about what renormalization means one
can calculate correlation functions. However no matter what one does, the
renormalized correlations do not obey the functional integral representation
i.e. the construction is a one-way street.} but no mathematical
status\footnote{Any method, even if it is entirely metaphoric, is useful as
long as it produces interesting new results. After 5 decades, there is however
the danger that it solidifies in our heads and becomes identified with QFT.}.
As euclidean representations they are far removed from modular localization
properties of section 3.2, hence it is not surprising that already in QM they
they are inferior to operator methods, less so in QFT where modular operator
algebra methods recently led to the first existence proofs for factorizing
d=1+1 models \cite{Lech} (which maybe viewed as the simplest field theoretical
analogs of intrgrable QM). Apart from superrenormalizable models in d=1+1
Lagrangian and functional methods have not attained a mathematical status
outside QM. As a result of their intuitive appeal, they are playing the role
of the great communicator between particle theorists with different
theoretical background. The writing down of an action for a particular model
is the common ground for starting a joint discourse and people with a good
conceptual understanding of QFT know how to use them as an intuitive launching
pad for specific conjectures.

Returning to the issue of spacetime embedding and restrictions, what is always
mathematically possible is the opposite of an embedding, namely a restriction
of higher dimensional QFT to a lower dimensional one; but even in this case
the problem of whether the restricted theory has reasonable physical
properties remains to be checked on a case to case basis. QFTs formally
associated with Lagrangians have the physically correct cardinality of phase
space degrees of freedom, but when one restricts such theory to a lower
spacetime dimension without a Lagrangian for the restricted theory, the naive
argument that this cardinality may be too high for accomodating a physically
viable QFT on the lower dimensional side has to be taken serious. This indeed
does render holographic projections \textit{physically} questionable, but
there is one notable exception in which the degrees of freedom adjusts
themselves dynamically to the lower dimension namely the holographic
projection from the bulk to its (causal or event) horizon.

This has the consequence that without knowing some feature of the bulk (in the
free case the mass of the theory), it is not possible to return with the full
information content from a horizon to the original bulk theory since e.g. the
knowledge of physical masses gets lost in the projection on a null-surface. In
all other cases, in particular in the case of the AdS$_{n}-$CFT$_{n-1}$
correspondence, one ends up with a mathematically existent, but physically
sick situation. A gigantic globalized community whose members produced more
than 6000 publications on this topic watches almost 20 years over this
conjecture but nobody within the community looked at the degrees of freedom
problem. The physical concequences of having too many phase space degree of
freedom are well-known: loss of causal propagation, occurrence of Hagedorn
temperature or worse).

As long as one thinks about QFT in terms of Lagrangian quantization, there is
no problem since classical relativistic Lagrangians (generalizations of
Maxwell's theory) have a causal relativistic Cauchy propagation which is
inherited by the QFT solution (if it exists), and the thermal physics based on
their quantization defines what we consider as a normal physical thermal
behavior. But outside the Lagrangian setting the validity of the causal shadow
property (the operator algebra in a spacetime region $\mathcal{O}$ is equal to
that in its causal shadow $\mathcal{O}^{\prime\prime}$) as well as normal
thermal behavior (thermal states for all temperatures $T>0$) has to be
established. If Jordan would have succeeded to formulate QFT without a
Lagrangian (in his terminology without "classical crutches", see next
section), he would have run into precisely this problem. The realization of
the significant difference between the cardinality in (zero density) QM
(finite number of phase space degrees of freedom per phase space cell) and QFT
(cardinality per cell is that of a compact or nuclear set \cite{Haag}) came
much later.

The Maldacena conjecture fails precisely on this issue i.e. a physical
AdS$_{5}$ theory becomes an unphysical CFT$_{4};~$this is a structural
property without exceptions. It does not help to start the other way around,
in this way the higher dimensional side would suffer from the opposite disease
of "physical anemia" i.e. a physically unacceptable lack of sufficient degrees
of freedom for finding point- or string-like generating fields of the local
akgebras in the AdS$_{5}$ spacetime. What is not excluded by this physical
No-Go theorem for having two corresponding \textit{physical} theories is the
possibility that the respective unphysical side may turn out to be
mathematically useful to learn something new about the physical side. After
all the the mathematical correspondence is a very radical spacetime
reprocessing of the same abstract quantum matter. Physics depends on both, the
given abstract quantum matter and its order in spacetime.

Returning to the dual model, the strange reinterpretation of a chiral
conformal theory as a dual model approximation of an S-matrix permits a vast
generalization which confirms the absurdity of its S-matrix interpretation.
The natural conceptual-mathematical setting of the duality relation has
nothing to do with particle physics of the S-matrix and its on-shell crossing
properties. Rather each conformal QFT, independent of its spacetime dimension,
is naturally associated with a dual model through its appropriately defined
Mellin transform \cite{Mack}. The sum over infinitely many intermediate poles
in the Mellin transform corresponds to the infinitely many terms in a global
conformal operator expansion. The residuum has precisely the desired form as
required by the dual model or to express it historically correct: the poles
and residua are what Veneziano found when (playing with gamma functions) he
came across the rules of the appropriately normalized Mellin transforms of a
conformal model but thinking that he found an approximate solution of
Mandelstam's crossing symmetric S-matrix program. The S-matrix and especially
the origin of its crossing properties was terra incognita at that time so that
any proposal which was mathematically consistent and fulfilled the at that
time accepted idea about crossing could pass as a (nonunitary) approximation
of an S-matrix with a given infinite particle multiplet.

Of course the elegant identification with Mellin tranforms of conformal
correlations is of a fairly recent vintage and hence the sense of having
stumbled upon something at least mathematically very deep by pursuing a
phenomenologically motivated program created a lot of excitement of having hit
an important structure of particle physics. The only additional requirement
which was not part of the original dual model, but rather entered through the
string theoretic interpretation, is that the \textit{inner symmetry space of
the conformal model should be identified with the spacetime} arena for the
action of a unitary representation of the Poincar\'{e} group. For anybody to
whom string theory has the appearance of being somewhat dodgy, this idea of
creating higher dimensional spacetime on the inner symmetry arena of a low
dimensional (chiral) QFT is probably the cause of that dodginess.

It is generally not possible to have a representation of a noncompact group as
inner symmetries (transformations on field components) since it is known from
higher dimensional QFT that the superselection theory requires internal
symmetries to be compact. However in d=1+1 the internal symmetry structure is
less restrictive and lo and behold, there is really an almost unique solution
of this physically unmotivated property, namely (up to a finite number of
M-theoretic modifications) the unique 10-dimensional "superstring", except
that \textit{the terminology "string" is misleading}\footnote{The reader
should be aware that whenever we continue to use this almost 5 decades old
terminology, it is in his historical and not in its physical meaning.} since
the infinite component wave function space and its second quantized operators
in Fock space have pointlike generators and form what may be called a
"pointlike dynamical infinite component field". The addition of "dynamical" is
to emphasize that these infinite component free theories are nontrivial in the
sense that their algebra contain more operators than those which transform
within each level of an infinite mass/spin tower. Without such level
connecting operators in the wave function space which determine the mass/spin
tower, the infinite component field would be uninteresting. In previous
attempts at dynamical infinite component theories the mass/spin spectrum was
expected to come from noncompact groups which generalize the Lorentz group
\cite{Tod} but the idea did not work and the project was abandoned.

In order to conclude that the quantization of the bilinearized Nambu-Goto
Lagrangian is pointlike generated one would not have to run through any
calculation; it would be sufficient to check that in the representation of the
Poincar\'{e} group resulting from that model there is no Wigner infinite spin
component, which for systems with a quadratic Lagrangian is the only way to
avoid stringlike covariant theories\footnote{In interacting theories there
could be algebraic string generators which lead to states which permit
pointlike generation (the Buchholz-Fredenhagen strings \cite{Bu-Fre})..}. The
explicit calculation is however useful in order to confirm that the
requirement of implementing a unitary positive energy representation
representation of the Poincar\'{e} group leads to pointlike dynamical infinite
component supersymmetric field in 10 dimensions. The interpretation of our
4-dimensional living space as coming from a dimensional reduction of this
space remains however in the eyes of the beholders.

But how is such a conceptual mistake about localization possible 40 years
after Jordan? Is it that physicists in string theory communities compute
correctly but fall behind previous generations if it comes to conceptual
problems or is perhaps the problem of localization a very subtle problem on
which one could easily slip? It is a bit of both. Despite a significant
difference between the Born-Newton-Wigner localization which has to be added
to QM for its interpretation, and the intrinsic modular localization of QFT
(the modern version of Jordan's causal localization), the issue of
localization has remained a rather neglected area of QT. Whereas the embedding
of a lower dimensional QM into a higher dimensional one and a Kaluza-Klein
dimensional reduction is trivially possible, the holistic aspect of modular
localization prevents the realization of this simple-minded idea. What happens
concretely in the case at hand is that the quantum mechanical chain of
one-dimensional oscillators has no choice to participate in the spacetime
localization of a higher dimensional "target". Rather the whole chain is
mapped into the internal mass/spin tower i.e. into an internal quantum
mechanical Hilbert space over one point which normally accommodates the spin
indices; it does not change the holistic localization structure of QFT. To
talk about such an internal "string" is certainly not what string theorists
mean and also would create terminological problems with situations of true
spacetime strings as e.g. electrically charged fields.

The conceptual confusion did not come out of nowhere; for many decades the
delusive terminology "relativistic QM" has been used when QFT was really
meant. There is no conceptual basis for objects which interact according to
Feynman like pictures involving splitting and recombining tubes. One of course
take the point of view that these rules can make sense by themselves even if
the result of the canonical quantization of the Nambu-Goto Lagrangian are
pointlike localized objects. But then one would have to prove (as Stueckelberg
did in the case of world lines) that the world tube recipe leads to amplitudes
which can be expressed \textit{in terms of states and operators}, the minimal
demand for something which is claimed to be of "quantum" nature. Functions and
recipes how to combine them are no replacement for a unitarization in terms of
operators and states. But this has been tried in vain for many decades as long
as string theory exists.

Another metaphoric idea which was already mentioned is the use of special
properties\footnote{Whereas inner symmetries in higher dimensionsn are
described in terms of compact groups, chiral theories permit in addition to
"rational" theories also irrational ones which are analogs of noncompact
symmetries (example; the continuously many charged represntation if chiral
current algebras).} of internal symmetries of chiral theories as carriers for
fundamental spacetime properties; this is nothing short of mysticism. The
rational way for understanding Witten's M-theory conjecture would simply
consist in forgetting the the spacetime metaphor and analyze these properties
at the place of their origin, namely as special properties of noncompact inner
symmetries of certain chiral current theories, but this would be less sexy.
But what the heck has the spacetime of Einstein's relativity to do with the
greater richness of "inner symmetries" which one meets in nonrational chiral models?

One could of course ignore these somewhat bizarre developments, but there is a
point of genuine preoccupation which results from the fact that, as explained
in the previous section, the important gauge theories do contain bona fide
string-localized interacting fields and may be affected by all these hardened
misunderstandings about string-localization localization which do not seem to
find an end. This weakening of pointlike field localization in those theories
is not directly visible in the infrared divergencies of global "on-shell"
observables and their replacement by inclusive cross sections in cases where
the latter can be defined as substitutes. Whereas string directions (points in
2+1 dimensional de Sitter space) appear \textit{explicitly} in charge-carrying
fields, their occurrence in global on-shell quantities is expected to be more
discrete in form of residual dependence on the string directions in the low
energy behavior of inclusive cross sections. In the standard gauge description
of Yang-Mills theories the infrared divergences are so strongly intermingled
with the ultraviolet renormalization that no renormalized correlations of
Lagrangian fields have been constructed. One hopes to sort out these problems
with a different formalism which takes care of the string-localization of the
vectorpotential (previous subsection) and hopefully leads to an understanding
of confinement sd the invisibility of certain strings.

The issue of string localization is even more important for zero mass higher
spin models e.g. the interactions of s=2 tensor potentials. Furthermore the
presently best way to understand renormalizable s=1 massive vectormeson
couplings to lower spin quantum matter is to start from e.g. QED with scalars
and spinors and force the scalars to undergo a Schwinger-Higgs screening. The
latter is similar to the Debeye screening of a Coulomb system in QM, but much
more radical in that the Debeye change of range of forces from long to short
range corresponds now to the "re-localization" of string-localized to
point-localized massive fields.

At no other time there has there been such little foundational progress in
particle theory as during the last 3 decades. This can be exemplified in terms
of changes how people thought about the gauge theoretic view of massive
vectormesons in renormalizable interactions with low spin quantum matter. The
Schwinger-Higgs screening mechanism of the 60/70s which has a perturbative
formulation starting from scalar QED is nowadays referred to the Higgs
mechanism with the "God particle" being the real particle which remains from
the screening of the complex charged field to a real neutral field. This is
more than a change of terminology since it leads to misinterpretation as a
spontaneous symmetry breaking, whereas in reality the symmetry disappeared
because the integral over the zero component of the current simply vanishes
whereas for spontaneous symmetry breaking it would diverges as a consequence
of the presence of a zero mass Goldstone particle. A more profound distinction
of what belongs to intrinsic properties of QFT and what is merely part of a
computational recipe in a Lagrangian setting would be helpful; unfortunately
textbooks do not contain discussions on these issues. The spirit of
"superstring theory" could perfectly survive its subject because its main
success is the formation of a cordon sanitaire against critical influences
around a community.

How much the corrective critical counterbalance has lost its influence on the
course of events can be seen by noting the comments of some famous
mathematicians when they address a mixed audience of mathematicians and
physicists. Among some praise about the depth of the mathematics-physics
encounter which finally has been reached thanks to string theory and its
derivatives, there is usually also an accolade of the Maldacena result on the
gravity-supersymmetric Yang-Mills theory correspondence. To anybody who knows
the subject it is evident that they don't know what they are talking about. To
get a feeling for the strangeness of this situation just imagine that at the
time of the discovery of QM and QFT in 1925 Hilbert and his colleagues from
the Goettingen mathematics department would have presented lectures at
seminars and conferences in which they would have praised the results of their
colleagues. They may have done this later without causing harm after a good
part of the new QT had been conceptually and mathematically established. was
established. But the Maldacena conjecture is almost 20 years after its vague
formulation even further away from a proof than before. It is a good thing to
start a study of a problem with an idea of what can be expected. However if
this is done by hook or by crook than it becomes an ideology In fact this is a
paradigmatic illustration to the kind of physics which grows in the
monocultures of globalized communities without a critical corrective.

\section{The 1929 Kharkov conference}

It is interesting to dedicate a separate section to the 1929 Kharkov
conference since this was the high point in the first phase of QFT which in a
way was also the culmination of Jordan's career in QFT. It was not only his
professional "swansong" in QFT before he, starting in the middle of the 30s,
increasingly isolated himself from the international community as the result
of his Nazi sympathies, it was also the swansong of Germany's leading role in
physics since a few years later German Jewish scientist who made significant
cultural and scientific contributions during imperial times and the Weimar
republic were forced by the Nazi government to leave their homeland, a brain
drain from which Germany never fully recovered.

At this 1929 conference in Kharkov\footnote{Landau, after his return from a
visit to Copenhagen, went to the university of Kharkov which for a short time
became the \textquotedblleft Mecca\textquotedblright\ of particle physics in
the USSR. The conference language at that time was still German} Jordan gave a
remarkable plenary talk \cite{Kharkov}. In a way it marks the culmination of
the first pioneering phase of QFT; but it already raised some of the questions
which were only taken up and partially answered almost 20 years later in the
second phase of QFT (i.e. in renormalized perturbation theory and its
application to gauge theory). In his talk Jordan reviews in a very profound
and at the same time simple fashion the revolutionary steps from the days of
matrix mechanics to the subsequent formulation of basis-independent abstract
operators (the transformation theory whose authorship he shares with Dirac and
London) and steers then right into the presentation of the most important and
characteristic of all properties which set QFT apart from QM: Commutation
Relations in agreement with causal locality and localization of states, as
well as the inexorably related vacuum polarization.

Already one year before in his Habilitationsschrift \cite{Habil} he identified
the two aspects of relativistic causality namely the statistical independence
for spacelike separations (Einstein causality, commutation of observables) and
the complete determination of events in timelike directions (the causal shadow
property) as playing a crucial role in the new quantum field theory. This
almost modern conceptual viewpoint of QFT was quite a shot away from the
"collection of oscillators" formulation in the Dreim\"{a}nnerarbeit where
Jordan's main motivation was to show that Einstein's photon contribution in
the statistical mechanics luctuations of a blach body radiation gas can also
be obtained from a fluctuation calculation on a system of a free photon in the
vacuum state. The more general project to generalize this approach to the de
Broglie matter waves led to the unified description of matter and light in the
form of QFT.

Jordan ends his 1929 conference talk by emphasizing that even with all the
progress already achieved and that expected to come in order to clarify some
remaining unsatisfactory features of gauge invariance (\textit{Die noch
bestehenden Unvollkommenheiten, betreffs Eichinvarianz, Integrationstechnik
usw., duerften bald erledigt sein)}, one still has to confront the following
problem: \textit{Man wird wohl in Zukunft den Aufbau in zwei getrennten
Schritten ganz vermeiden muessen, und in einem Zuge, ohne
klassisch-korrespondenzmaessige Kruecken, eine reine Quantentheorie der
Elektrizitaet zu formulieren versuchen. Aber das ist Zukunftsmusik.} (In the
future one perhaps will have to avoid the construction in two separated steps
and rather approach the problem of formulating a pure quantum theory of
electromagnetism (a pure QED) in one swoop, without the crutches of classical
correspondences. But this is part of a future tune.)

He returns on this point several times, using slightly different formulations
(....\textit{muss aus sich selbst heraus neue Wege finden...) }for a plea
towards a future autonomous formulation of QFT which does not have to take
recourse to quantization which requires starting with an (at least imagined)
classical analog.

These statements are even more remarkable, since they come from the
protagonist of field quantization only four years after this discovery. A
similar curious flare up of an important idea ahead of its time was the
attempt to generalize QED which was rediscovered in the proceedings of a 1939
conference in Warsaw. It contained a talk by Oskar Klein, Jordan's
collaborator from the Copenhagen time, grappling with the intricacies of
nonabelian gauge theories and their possible applications to particle physics.
It seems that the knowledge about QFT in pre-war Europe was more advanced then
hitherto thought.

Jordan%
\'{}%
s critical attitude towards his own brain child of wave-field quantization has
the same philosophical origin as his antagonism to Dirac%
\'{}%
s particle approach. Being a positivist, he had no problems with the physical
non-existence of a classical structure to be quantized, as long as the
quantization was consistent and gave the experimentally verifiable behavior of
quantum matter. To him quantization from a uniform setting of classical matter
waves was preferable to Dirac's dual program of wave quantization of light and
particle quantization for massive matter.

What bothered Jordan about field quantization was the apparent necessity of a
parallelism to classical physics which is inherent in the very procedure of
quantization, be it that of particles or that of fields. Clearly a fundamental
physical theory should stand on its own feet and explain the classical
behavior in certain limits, and not the other way around, as in the various
quantization approaches. Neither Dirac's nor Jordan's own approach was able to
meet this plea for an autonomous approach to QT.

Often the joint work of Jordan and Klein \cite{J-K} is cited as the inaugural
paper on field quantization. This is an easily corrected conceptual
misunderstanding. The Jordan-Klein paper is an exposition of the formalism of
second quantization in the Fock space setting of multiparticle QM. This has
nothing to do with the physical content of QFT, it is just a condensed
notation for writing the n-particle Schr\"{o}dinger equations, which uniquely
follow from a two-particle interaction, in terms of one Hamiltonian or one
field equation\footnote{This Fock space formalism turns out to be quite useful
in condensed (nonrelativistic) natter physics at finite density.} which does
the job for all particle number n in one sweep. Interacting QFT acts in a
Fockspace, but there is no two- or higher- particle relativistic equation from
which it arises through "second quantization", in fact apart from free fields
which arise from Wigner one-particle representations in a functorial way,
there is no setting from which QFT arises by second quantization; the latter
is simply incompatible with causal localization-preserving interactions.

The only structure which resembles the Jordan-Klein Schr\"{o}dinger field
formalism occurs in the absence of interactions when the one-particle spaces
may be classified in terms of Wigner's representations of the Poincar\'{e}
group. In that case one may define a functor which leads from the modular
localized subspaces of the Wigner representation space to the net of local von
Neumann algebras as the "second quantization functor" as explained in the
previous section. But such a terminology would be limited to interaction-free theories.

In retrospect it is clear that an autonomous approach to QFT had no chance
before the formalism of interacting field quantization was fully developed to
the level of the conceptually sophisticated renormalized perturbation theory,
as in the work of Tomonaga, Feynman, Schwinger and Dyson in the end of the
40's. Only after this seminal work, the mathematical status of the inherent
singular nature of pointlike fields became gradually understood, and arguments
which permit to avoid infinities in intermediate computational steps were
proposed. An autonomous formulation of QFT which is capable to classify and to
construct models has been in the making since the beginning of the 60s; it had
a slow start but gained increasingly steam during the last two decades. What
is referred to as LQP \cite{Haag} or AQFT is a set of physically motivated and
mathematically well-formulated requirements which allow to derive interesting
general consequences. In d=1+1 one knows additional structures which permit a
partial classification and construction of models. The progress obtained
during recent years nourishes the hope that a complete knowledge of autonomous
QFT including a classification of models and a proof of their mathematical
existence is possible in the realistic case of d=1+3. None of these recent
attempts in LQP can be traced back to Jordan's 1929 plea for a QFT; the time
lapse is too large and world war II has interrupted a continuous development.
Nevertheless it is encouraging to know that the first pleas for an autonomous
understanding of QFT are even older than renormalization theory, deep ideas
have their harbingers before they take roots. Some more remarks about the post
Jordan development of these concepts will appear later.

Jordan's expectation about a rapid understanding of the \textquotedblleft
imperfections of gauge theories\textquotedblright\ at the time of his Kharkov
talk may have been a bit optimistic since the Gupta-Bleuler formalism (much
later followed by the more general BRST approach) only appeared 20 years
later. But it is interesting to note that for Jordan gauge theory was an
important issue already in 1929. His view of gauge theories was, as that of
the majority of contemporary "applied" quantum field theorists, just taken
over from the classical use of vectorpotentials where they are an
indispensable computational tool (example: Lienhard-Wiechert potentials). It
is questionable that he had a foundational QFT view as that explained in
section 3.2 which sees quantum gauge theory as the consequence of resolving a
fundamental clash between the quantum Hilbert space structure with the
existence of pointlike vectorpotentials which either requires to abandon the
Hilbert space structure in order to save an initially unphysical but
technically well-known perturbative formalism, or to confront dealing with
stringlike potentials as in subsection 3.2.

Finally one should add one more remark which shows that Jordan and Dirac had a
very similar taste for what both considered relevant problems. In 1935 Dirac
presented a beautiful geometric argument which establishes that it is possible
to introduce a magnetic monopole into quantum electrodynamics as long as the
strength of the monopole times the value of the electric charge fulfill a
quantization law. In the same year Jordan came up with an interesting very
different algebraic argument for the magnetic monopole quantization which he
based on the algebraic structure of bilinear gauge invariants \cite{Mono}. He
published a short note in Zeitschrift fuer Physik at a time when its
international reputation had suffered from the deteriorating political
situation (see first footnote). Hence it is not very surprising that this
paper was hardly noticed. The independent re-discovery of its main content
(including Jordan's tetrahedron argument) by R. Jackiw \cite{Jack} certainly
shows that the problem remained non-trivial and interesting almost 70 years
later. The problem of how these kind of arguments have to be amended in order
to take account of renormalization has according to my best knowledge not been
satisfactorily answered. Jordan's monopole paper is one in series of
publications in which gauge invariance based on exponential functions of
line-integrals over vectorpotentials play a central role (see next section).
It is clear that what Jordan really wanted to achieve was a gauge invariant
formulation of the basic dynamical equations of quantum electrodynamics. This
he did not achieve, but attempts 30 years later, in particular by Stanley
Mandelstam, also failed. Although gauge theory aims at gauge invariant
observables and as such fulfills Heisenberg's dictum that the use of
non-observables must be avoided since, it does use unobservable negative
metric ghosts in intermediate steps. Their use is reminiscent of a "catalyzer"
in a chemical reaction which can be removed after having done its job. Whereas
their mode of action in chemistry is well-understood, there is no fundamental
understanding why one needs such ghost catalyzers for the description of
interactions involving higher spin equations. Their use is obviously helpful
for the construction of gauge invariant local observables, but the necessarily
nonlocal physical charge-carrying fields remain outside the standard gauge formalism.

Since Jordan left the area of QFT in the middle of the 30s and became
disconnected from the important post-war discoveries, we do not know how he
would have reacted to the amazing progress brought about by renormalization
theory. Probably, as most of the leading particle theorist of the thirties,
Jordan would have expected that only a revolution in the conceptual
foundations could save QFT as a theory of relativistic particles. In that case
he would have been deeply surprised that the post war progress resulted from a
careful conceptual distinction between formal and physical (observed)
Lagrangian parameters, as well as by a systematic replacement of the old
quantum mechanic-inspired formalism by a more intrinsic field theoretic
relativistically covariant setting and a lot of hard computations. In the
words of Weinberg the success was achieved in a rather conservative manner
\cite{Wein}.

In fact renormalization theory may have been too conservative in order to fit
Jordan's radical expectations; in strange contrast to his reactionary
political stance, in physics Jordan was a visionary revolutionary.

\subsection{Local Quantum Physics and Jordan's ideas about a future QFT}

30 years after Jordan's state of the situation presentation in Kharkov a
conceptional renewal of QFT was set in place which nowdays is known as local
quantum physics (LQP) or algebraic QFT (AQFT). Its immediate predecessors were
Wightman's formulation of QFT, which gave rise to the first mathematical
treatments \cite{Wight}, and the LSZ setting linking asymptotes of fields with
particles \cite{Haag}. But the project which fits best Jordan's quest for an
intrinsic formulation without any quantization ties to a classical parallelism
is that which Haag outlined for the first time at a conference in Lille in
1957 \cite{Lille}. By that time Jordan had become the unsung hero of QFT and
there was certainly no historical connection of the new attempts to get a more
intrinsic access to local quantum physics (LQP = QFT with the emphasis on the
causal locality principle) with the old prophetic pronouncements at the
Kharkov conference.

Up to the 60s a lot of things had happened in QFT. There was a much better
understanding of the true nature of quantum fields. In contradistinction to
classical fields and second quantized Schr\"{o}dinger fields (as von Neumann
showed it is not necessary to know the delta function in QM) the new
distribution theory of Laurent Schwartz of the 50s was essential in order to
understand the conceptual position of quantum fields within QT. The so called
ultraviolet divergencies of QFT which led almost to its abandonment were
partly cause by treating fields as operators and not as operator-valued
distributions. In the 60s it became clear in the work by Epstein and Glaser
\cite{E-G} that a perturbative classification and construction of
renormalizable models can be given in the setting of time-orderd correlation
functions of fields in terms of the short distance scaling degrees of local
polynomials in the free fields describing interactions. If the short distance
power counting leads to a degree less than 4 (in d=1+3), than the perturbative
series can be written in terms of a finite number of interaction parameters
independent of the order and the theory is called renormalizable. Different
from standard quantum mechanical inspired approaches, which try to renormalize
infinities which were the result of incorrectly handled singular aspects of
pointlike fields (operator-valued distributions) and their composites, the E-G
method iterates the zero order input according to the locality principle and
stays in a multi-particle Fock space. This was the kind of perturbative
formulation which fits best Haag's impressive design and Jordan's dream.

An important aspect of that construction was the explicit knowledge of the
equivalence class of all fields which are Einstein causal relative to the free
field. This local equivalence class consisted of the pointlike Wick
polynomials of a free field. This and other structural results coming from
causal localization (modular localization in the more recent mathematical
setting) permitted to base perturbation theory on physical principles rather
than on approximating concrete operators as in QM. But the results were of
course the same as that of the cutoff or regularization methods, the only
reason for mentioning the finite methods is that some people (especially
string theorists) believe that QFT is beset my unavoidable ultraviolet
problems whereas the real problem of perturbation theory is the number of
coupling parameter which limit the predictive power and which render
nonrenormalizable interactions useless. Later it became clear that
renormalized perturbative series never converges. This implies that
perturbation theory has no explanatory power concerning the existence of these
models; the best one can hope for is that there is a notion of asymptotic
convergence in the limit of vanishing couplings which could be behind the
often spectacular experimental agreement in QED and weak interactions.

The structure of local equivalence class of (infinitely many) pointlike fields
suggests that fields in such a class have, different from classical fields, no
individual distinction from each other apart from their charge; locally
equivalent fields which carry the same superselected charge create the same
particles and lead to the same scattering theory (interpolating fields for the
same particle scattering). Hence the use of different fields in such a class
in a fixed QFT is like the deployment of different systems of coordinates in
geometry. Haag build his LQP on the idea that it is not the individual field
but a spacetime indexed net of operator algebras. In this way the individual
fields disappeared and only the concept of charge remained with the algebras;
similar to coordinates in geometry they now played the role of generators of
algebras of which there were infinitely many. With the spacetime-indexed net
of algebras as defining a QFT, the confusing plurality of QFT disappeared and
instead the question arose whether this setting describes all physical
phenomena which were previously described in terms of individual fields. The
answer to this question is positive if one attributes a preferential status to
conserved currents which according to Noether's theorem already had a
distinguished status in the classical theory. One can show that also in the
algebraic setting conserved currents can be constructed from a (localized
version) of symmetries of the QFT and this construction can be generalized to
spontaneously broken symmetries.

It is much more difficult to convince quantum field theorists that the
intrinsic local operator algebra approach is the most suitable in order to
take on the remaining difficult problems of particle theory whereas they
except immediately that in mathematics the coordinate-independent intrinsic
formulation of geometry is the most adequate one. Often those particle
physicists who have internalized the intrinsic formulation on the mathematical
side (fibre bundles, cohomology,...), turn out to be the fiercest defenders of
the status quo if it comes to the conceptual-mathematical setting of QFT.

For the first three decades the intrinsic setting of algebraic QFT has been
predominantly applied to the solution of deep structural problems as e.g. the
quantum origin of the notion of internal symmetries which for the first time
entered nuclear physics through Heisenberg's isospin. This was not part of
classical physics (although often physicists read such properties back into
the classical setting) so it was natural to ask the question about its origin.
In Haag's setting of local quantum physics the question took on the more
specific form: what is the role of locality in de-mystifying internal
symmetries? The answer was that the localizable representations of an
observable net (the localizable superselection sectors) have the structure of
a dual of a compact group and the best way to present this situation consists
in defining a field algebra consisting of Bose and Fermi fields on which this
internal symmetry group acts such that the original charge neutral observable
algebra re-emerges as the fixpoint algebra \cite{Haag}\cite{Do-Ro}. Being a
structural theorem, this does not distinguish a particular compact group (in
fact it can be shown that all compact groups can occur). Somewhat exaggerated
but logically correct one may say that if groups were not discovered in other
purely mathematical contexts way back, the quantum locality principle would
have also spotted this structure.

QFT is different from any other theory in physics, including QM, in that one
has not been able to come up with nontrivial example involving interactions.
This is a problem Jordan must have been aware of because what he too
optimistically called "neutrino theory of light" can be considered as a result
of searching in d=1+1 for well-defined nontrivial models. As mentioned
renormalized perturbation theory did not solve the problem of finding
interacting models either since there are rigorous statements which show that
the there are no circumstances under which the perturbative series can
converge. LQP only postponed but did not suppress the longing for an existence
demonstration of a nontrivial illustration for interacting QFT. Finally,
starting in the 90s, some ideas which were endogenous to LQP appeared
\cite{AOP}. They led to successful existence proofs for QFT from the class of
factorizing models \cite{Lech}.

These existence proofs follow a different logic from that suggested by the old
functional analytic Glimm-Jaffe arguments which were limited to theories with
the same short distance behavior as free ones (superrenormalizable), such
couplings only exist in d=1+1; it is also totally different from what a
functional integral representation would suggest. The idea is to start as far
as possible from pointlike fields i.e. with objects which have the least
amount of vacuum polarization i.e. with the coarsest localized objects because
they have the least amount of vacuum polarization. These are generating
operators of wedge localized algebras. The mentioned class of models is
distinguished by rather simple properties of wedge generators which lend
themselves to a classification. The compact localized algebras which have the
full vacuum polarization and whose generators are the pointlike localized
quantum fields are then obtained by intersecting wedge algebras and the hard
part of the existence proof consists in showing that these intersections are
nontrivial and act on the vacuum in the standard manner.

Even the perturbative approach, which has the closest contact with
measurements and phenomenological ideas, is incomplete. More precisely it is a
closed subject only in case of pointlike interactions but not in case of gauge
theories and interactions involving massless higher spin representations. In
that case one needs a generalization of the Epstein Glaser approach to
stringlike free fields since this is the only possibility to understand the
origin of string localization charged matter fields whose's delocalization
comes about through interactions with string-localized free vector potentials.

It is interesting to note that the factorizing models are classified by
elastic S-matrices. Since the cardinality of factorizing S-matrices is much
bigger than that of interacting Lagrangians, most existing factorizing QFTs
have no Lagrangian name i.e. the Lagrangian quantization setting does not
exhaust the possibilities of models of QFT. This illustrates the validity of
Jordan's point that the use of classical crutches in the form of Lagrangian
quantization may turn out to be too narrow.

\section{"Bosonization" instead of "neutrino theory of light"}

Starting around 1935 Jordan began to publish a series of papers under the
title "On the neutrino theory of light" \cite{Jo}. The idea that photons may
be bound states of $\nu$-$\bar{\nu}$ was not entirely new since de Broglie had
vaguely formulated a similar thought but without presenting an argument. The
statistics of particles in terms of commutation relations of fields was still
an unaccustomed subject and therefore problems which nowadays are considered
as part of kinematics to be done away with in a few lines, at that time filled
the main part of a paper. There was the general belief from QM that it is
sufficient to illustrate an idea in a low-dimensional QFT; we know thanks to
Wigner's work on Poincar\'{e} group representation theory of particles that
this is incorrect. This explains to some extend why Jordan's contemporaries
had no problem with the fact that he took a two-dimensional model instead of
arguing in the realistic setting of 4-dimensional spacetime. Jordan started
from a two-dimensional massless Weyl fermion, his neutrino model, from which
he formed a bosonic current and its potential which was then his
two-dimensional analog of the photon field.

His model amounts to what since the 70s is called \textit{bosonization} and
\textit{fermionization }refers to\textit{ }its inverse; both procedures only
work in two dimensions and shed no light on a higher dimensional physical
neutrino theory of light. But although Jordan was wrong in his central claim,
he discovered at least the mathematics of an interesting new structure for
which however there was no demand in QFT prior to the 60s. Before the
presentation of Jordan's model in more detail, some additional remarks on the
state of QFT at the time of Jordan's participation are in order.

Using a modern notation and terminology his main points in \cite{neutrinoII}
become more accessible. Starting from a $u=t+x$ chiral free fermion $\psi(u)$
one may define the u-component of a chiral current%
\begin{align}
\psi(u)  &  =\frac{1}{(2\pi)^{1/2}}\int_{0}^{\infty}dp(e^{ipu}a(p)+e^{-ipu}%
b^{\ast}(p))\label{current}\\
j(u)  &  =:\psi^{\ast}(u)\psi(u):~=~\partial_{u}V(u),~~u=t+x\nonumber
\end{align}
The u-lightray component of the spinor has a $v=t-x$ counterpart which is not
needed here, but would be required for the massive spinor which depends on
both $u$ and $v$ (for which however the commutation properties are less
simple). The double dot denotes as usual the ordering in which all
annihilation parts appear in the right of the creation operators (with a sign
factor for each fermion commutation). The only commutation relation one has to
know in order to compute all the others is that between (Wigner) momentum
space creation and annihilation operators are the standard ones $\left\{
a(p)a^{\ast}(p^{\prime})\right\}  =\delta(p-p^{\prime})$ and similarly for the
$b.$

The surprise is the result of the $j(u)$ commutator. On would naively think
that the commutator of an expression which itself is bilinear in fermions
would contain in addition to a c-number term (complete contraction) also a
bilinear term with one contraction; but against naive expectation this
operator term (which would be present in higher dimensions) is absent for
2-dimensional massless fermions \cite{Klaiber}. In fact this is the only case
where this simplification occurs, in all other cases the bilinear current of a
Fermi field does not fulfill canonical commutation relations. This tricky
aspect which was which was correctly handled by Jordan was not understood by
his contemporaries, in particular it brought him into a conflict with Fock
\cite{Fock}. In the abstract of the above cited paper Jordan vigorously (and
correctly) refutes Fock's critique. Nobody at that time seemed to have found
it problematic to draw physical conclusions from d=1+1 zero mass "neutrinos"
and "photons" about the real d=1+3 situation; what was not accepted was that
the the current from a free massless spinor in d=1+1 is a canonical field,
just opposite from what we know nowadays!

It is worthwhile to write the form of the current commutation relation%
\begin{equation}
\left[  j(u),j(u^{\prime})\right]  =c\delta^{\prime}(u-u^{\prime})
\end{equation}
Jordan performed his calculation in the filled Dirac sea i.e. in the charge
symmetric prescription; the use of the \textit{hole theory} would have caused
a serious confusion in particular in such calculations. This derivative term
in a commutation relation was equivalent to the presence of the so-called
Schwinger terms of 1959 \cite{Schwi} i.e. derivative of delta functions in the
mixed space-time components of currents in any dimension.

The cited paper of Jordan also treats the inversion of bosonization, namely
the re-fermionization starting from the potentials $V(u)$%
\begin{align}
\Psi(u,\alpha)  &  \equiv e^{i\alpha V(u)}=e^{i\alpha\int_{-\infty}^{u}%
j(u)du}\label{fer}\\
\left[  j(u),\Psi(u^{\prime},\alpha)\right]   &  =\alpha\delta(u-u^{\prime
})\Psi(u^{\prime},\alpha)\nonumber
\end{align}
The case $\alpha=1$ which leads back to the zero mass canonical fermion with
scale dimension $1/2.~$From a modern point of view these fields are
charge-carrying fields of charge $\alpha$ associated with the current operator
$j(u).$ Each charge defines a superselection sector i.e. there are
continuously many. The $\Psi$ are fields which turn out to be conformal
covariant and therefore possess a scale dimension which is proportional to
$\alpha^{2}$ (the proportionality constant depends on the normalization of V).
Together with the anomalous scale dimension the charge $\alpha$ determines
also the anomalous (conformal) spin and through it the statistics which turns
out to be "anyonic" i.e. that associated with an abelian braid group
representation. There is one value of $\alpha$ for which the conformal spin is
1/2 and the anyonic commutation relation becomes fermionic; this is the value
which Jordan used for re-fermionization and for which he showed that the
$\Psi$ coalesces with the original fermion $\psi.$ Hence starting from a
chiral Fermion $\psi$ one passes to a chiral current $j$ which turns out to be
the derivative of an (infrared divergent) free field $j(u)=\partial_{u}V(u)$
which in turn through (\ref{fer}) leads back to a free Fermi field
$\Psi(u,\alpha=1)$ in a different veil. This passage is nowadays referred to
as bosonization/fermionization. It is limited to d=1+1 and in the above form
also to the chiral (the u,v lightray) components of a Fermion or a current.

The model is identical to that which Jordan used in the Dreimaennerarbeit to
support Einstein's fluctuation formula namely the solution of the wave
equation%
\begin{align}
&  \partial_{\mu}\partial^{\mu}\Phi(t,x)=0,~\Phi(t,x)=V(u)+V(v),~j(u)=\partial
_{u}V(u),~j(v)=\partial_{v}V(v)\label{chiral}\\
&  ~~\ T(u)=:j^{2}(u):~~~T(v)=:j^{2}(v):~~~~~~~T(x,t)=T(u)+T(v)\nonumber
\end{align}
Here the u and v dependent objects are independent singular operators and the
$T~$denote the lightray components of the energy momentum tensor. Instead of
considering these objects on their full u,v lightrays where they possess their
complete 3-parametric Moebius spacetime symmetry on each lightray, Jordan
quantized the system in a an interval of length L and studied the
T-fluctuation in a smaller interval I. The L interval quantization removes the
infrared divergence of the $V(u),$ although this would not have been necessary
since only the infrared well-behaved $j$ and $T$ enter Jordan's energy
fluctuation calculation. But quantum physicists (not only at Jordan's time)
preferred to deal with sums rather than integrals even if it meant loosing
helpful symmetries and shifting problems to other places. As a side remark,
the word string in Jordan's contribution is synonymous with a one-dimensional
periodic quantization box and has nothing to do with string-localization or
string theory since the fluctuation computation can be formulated in terms of
the pointlike localized current operator $j(u)$. The fluctuation region is an
interval of size $I$ inside the string interval of length $L$.

The modern derivation of Einstein's fluctuation formula would consist in
computing the localization energy associated to the interval I on the full u
or v lightray. In doing this one can bypass quantum mechanical arguments and
use concepts from modular localization which are intrinsic to QFT and have no
counterpart in QM. Since the restriction of the vacuum state to the operator
algebra of an interval $I$ is known to be a singular\footnote{The
thermodynamic limite state of an increasing sequence of Gibbs states for
V$\rightarrow\infty$ is the only KMS state in QM. It leads to finite thermal
correlation functions but admits no description as a denslity state (the
partition function diverges).} thermal KMS state with the generator of the $I$
preserving dilation being the "modular Hamiltonian", one has to approximate
this state (as in the case of the thermodynamic limit) by a sequence of Gibbs
states. Modular theory provides a canonical way for doing this: the split
property \cite{interface}. But although the modular localization theory
assures the existence of these Gibbs density states which converge against a
singular KMS state, there is yet no easy calculational approach for the
sequence of Gibbs like density matrix states. This step of approximating the
singular KMS state for the restriction of the vacuum to the sharp localized
interval by a sequence of \ density matrix states which are sharp- in I but
fuzzy-localized inside the $\varepsilon$-collor at the end points corresponds
precisely to Heisenberg's "fuzzy boundary" \cite{Du-Ja} where $\varepsilon
\rightarrow0$ corresponds to the return to the sharp boundary (which in chiral
models causes a logarithmic divergence but in higher dimensions contains
additional inverse powers \cite{BMS}). The localization energy and its square
fluctuation has not been calculated along these line, only the somewhat
simpler dimensionless localization entropy has been derived. In the
one-dimensional case of QFT on the lightray there is an exact relation between
the standard heat bath state associated to the tranlative Hamiltonian and the
thermal aspects caused by modular localization in an interval $I$ (the inverse
Unruh effect \cite{interface}\cite{BMS}). One expects that the leading
behavior of the energy fluctuation can be derived similarly. Since Einstein
uses rather general properties of thermal averages and the density matrix
state from the restriction of the vacuum to the operators localized in the
interval $I$ with fuzzy boundaries has the form of thermal state Gibbs state,
where the role of the heat bath temperature kT in Einstein's case is played by
the "geometric" modular temperature \cite{BMS}.

The main reason for mentioning these ideas based on modular localization in
connection with Jordan's quantum mechanically inspired calculation and its
recent improvement \cite{Du-Ja} is to emphasize that, although QFT was
discovered in 1925, the cutting of its umbilical cord to QM and the discovery
of its conceptual and calculational autonomy was a long-lasting process.
Without understanding QFT from its intrinsic conceptual modular localization
structure, which includes in particular the thermal aspects of QFT
localization, the relation between Jordan's calculation in the vacuum state
and Einstein's thermal energy fluctuation in a heat bath state would have
remained a plausible analog but surrounded by an air of mystery which no
quantum mechanical improvement can fully remove. A restriction of the vacuum
state in second quantized Schr\"{o}dinger quantum mechanics to an interval
would just remain a vacuum on that localized tensor algebra and not manifest
itself as a singular KMS state; it is not the second quantized field formalism
but rather the aspect of modular localization which distinguishes between QM
and QFT see \cite{interface}. These new localization-based concepts do not
only remove the loose conceptual ends of the past, but they also have led
recently (for the first time in the history of QFT!) to the rigorous
construction of certain two-dimensional models \cite{AOP}\cite{Lech} with
methods which are far removed from the shared common stock of (Lagrangian,
functional integral) quantization methods of QM and QFT. It seems that the
heavy investment into Haag's LQP \cite{Haag} project dedicated to the
discovery of the foundational properties of QFT, which started in the early
60s and gained pace in the last 3 decades, is now bearing its first fruits and
begins to revolutionize QFT in the direction of its autonomous understanding.

Returning to the work on bosonization/fermionization; what one does not find
in Jordan's papers \cite{neutrinoII} are calculations of states or correlation
functions. With QFT at that time still being tied down to the formalism of QM
it is doubtful that anybody at that time would have had the conceptual
resources one needs to calculate the correlation functions of the $\alpha
$-charged fields
\begin{equation}
\left\langle \Psi(u_{1},\alpha_{1})\Psi(u_{2},\alpha_{2})...\Psi(u_{n}%
,\alpha_{n})\right\rangle \neq0\text{ }only~for~\sum_{i=1}^{n}\alpha_{i}=0
\label{cor}%
\end{equation}
and extract the charge superselection rules from the infrared behavior of the
$V(u,\alpha).$ This certainly cannot be understood in terms of quantum
mechanical computational rules. A more demanding method based on the DHR
(Doplicher-Haag-Roberts) superselection method of algebraic QFT which is a
consequence of modular localization can be found in \cite{BMT}.

As already mentioned, at the beginning of QT there was a widespread belief
that one only needs to present an illustrative mathematically controlled model
in two-dimensional spacetime dimension, its validity in d=1+3 would then
follow by analogy. Whereas this is true for most problems of QM, this is not
so in QFT. The properties of relativistic particles are dependent on the
spacetime dimensions; there is no good analog of photons and neutrinos in
d=1+1. As mentioned nobody at the time criticized Jordan's two-dimensional
"neutrino theory of photons" for its lack of validity in the realistic case.
Only with the arrival of Wigner's intrinsic representation theoretical
approach to particles this dimensional dependence begun to attract notice. But
Jordan's contemporaries were deeply suspicious of the d=1+1 bosonization of a
Fermion current; as mentioned Fock wrote a counter paper \cite{Fock} in which
he claimed that Jordan made a computational mistake.

There was the feeling that with the neutrino theory of light Jordan had let
his imagination go overboard. \ Hence a little rubdown was in store. It came
as a carnavalesque "Spottlied" with the following text (the melody is that of
Mack the Knife) \cite{Pais}:

\textquotedblleft Und Herr Jordan
\ \ \ \ \ \ \ \ \ \ \ \ \ \ \ \ \ \ \ \ \ \ \ \ \ \ \textquotedblleft Mr. Jordan

Nimmt Neutrinos \ \ \ \ \ \ \ \ \ \ \ \ \ \ \ \ \ \ \ \ \ \ \ \ \ \ \ takes neutrinos

Und daraus baut \ \ \ \ \ \ \ \ \ \ \ \ \ \ \ \ \ \ \ \ \ \ \ \ \ \ \ \ and
from those he

Er das Licht
\ \ \ \ \ \ \ \ \ \ \ \ \ \ \ \ \ \ \ \ \ \ \ \ \ \ \ \ \ \ \ \ \ \ \ builds
the light.

Und sie fahren
\ \ \ \ \ \ \ \ \ \ \ \ \ \ \ \ \ \ \ \ \ \ \ \ \ \ \ \ \ \ \ \ And in pairs they

Stets in Paaren
\ \ \ \ \ \ \ \ \ \ \ \ \ \ \ \ \ \ \ \ \ \ \ \ \ \ \ \ \ \ \ always travel.

Ein Neutrino sieht man nicht.\textquotedblright\ \ \ \ \ \ \ \ \ One
neutrino's out of sight\textquotedblright.

One would suppose that this rather good humored song was presented at the end
of a conference or during a conference dinner, but Pais does not comment on
this. Insofar as the mock song refers to the misleading title it is
unintentionally correct, since they thought that the error was in the
presentation of the bosonization/fermionization idea in which the paper was
not only correct but even far ahead of its time! The mistake Jordan and his
taunters made was to take for granted that each phenomenon in low dimension
permits an extension to higher dimension; this is true in QM but not in QFT.

Our modern viewpoint on this issue is that although the photon cannot be
viewed as a $\nu$-$\bar{\nu}$ bound state, an interacting QFT is forced
(essentially by its locality principle) to follow the local quantum physical
adaptation of "Murphy's law" \cite{interface}: what is not forbidden (by
superselection rules) to couple does couple!

Applied to the problem at hand this means among other things%
\begin{equation}
\left\langle 0\left\vert F_{\mu\nu}(0)\right\vert p,\bar{p}^{\prime
}\right\rangle ^{in}\neq0
\end{equation}
In words: the formfactor of the photon field between a $\nu$-$\bar{\nu}$ state
with momenta $p$ and $p\prime$and the vacuum is nonzero (but as a result of
the presence of week and electromagnetic interaction it is extremely tiny).
The label in/out on particle states of formfactors is important because the
connection between particles and fields in the presence of interactions was
not at all understood at the time of Jordan but starting from the 60s we know
that interacting QFT has no particle at finite times, they only have an
asymptotic reality for infinite times in the sense of scattering theory. This
is sufficient to secure their existence as states in the physical Hilbert
space, but not for generating particles in compact spacetime regions. One
consequence of what has been metaphorically referred to as "Murphy's law" of
QFT is \textit{nuclear democracy} namely that the quantum mechanical hierarchy
between elementary particles and bound states disappears in QFT; the only
remaining hierarchy is that between fundamental and fused "charges". The
localization structure of the vacuum representation of the observable algebra
contain the informations which are needed in order to construct all non-vacuum
superselection sectors and combine them into a field algebra on which a
symmetry algebra acts in such a way that the observables re-emerge as the
fixpoint algebra under internal symmetry transformation \cite{D-R}.

At this point it may be helpful to take another short break and remind the
reader of the change of content of meaning of the term "QFT" during the
passage of time. Within the first two decades it was often used in the sense
\textit{QM of systems of infinite degrees of freedom}, in particular if those
infinite degrees of freedom result from Fourier decomposition of
operator-valued space(time) dependent fields. However the mere change of
formalism from a two particle Schr\"{o}dinger equation to a multiparticle
formalism as done in the work of Jordan and Klein \cite{J-K} would nowadays
not pass as a genuine illustration of the spirit of QFT, at best it serves to
illustrate some aspects of its formalism. Already at the time of Jordan the
"second quantization formalism" received mocking comments from some quarters
as: "Why quantize something a second time which was already quantum". A more
useful well-known comment by Edward Nelson settled this issue once and for all
by stating: "second quantization is a functor and quantization is an art".

An intrinsic characterization of QFT is based on (physically) causal or
(mathematically) on modular localization. Here "causal" refers to the kind of
relativistic causality of finite velocity Cauchy propagation for hyperbolic
equations. This can be formulated in an intrinsic way which does not refer to
quantization. The characteristic phenomenon of vacuum polarization as
discovered by Heisenberg in studying the relation of conserved currents and
their charges in free field theories \cite{Hei} and in the presence of
interactions in the work of Furry and Oppenheimer\footnote{In the presence of
interactions all fields, elementary and composites, applied to the vacuum
create in addition to the desired particle infinitely many
particle/antiparticle pairs (polarization cloud).} \cite{Fu-Op} is the
characteristic phenomenon of causal localization which has no counterpart in
the localization arising from a position operator in QM. An analog to vacuum
polarization (particle-hole pairs) is also encountered in case a second
quantized formalism is used in a ground state of finite density (the Fermi
surface); this explains why the formalism of QFT in certain cases has a useful
analog in solid state physics.

What sets QM and QFT apart in the most dramatic way are their totally
different localization concepts \cite{interface}. As a result the local
operator algebra in QM defined in terms of the second quantized
Schr\"{o}dinger fields are of the same type as their global counterpart namely
type I$_{\infty}$ von Neumann factor algebras; the local algebras
$\mathcal{A(O})$ of QFT in causally complete regions are radically different
namely hyperfinite type III$_{1}$ factors$,~$shortly called "monads" since
they are all isomorphic to such a monad in \cite{interface}). As a consequence
there is no tensor factorization between a localized operator algebra and its
commutant which is localized in the spacelike complement, even though these
two algebras commute! From this follows that there is no notion of
entanglement in this situation; i.e. the point of departure of quantum
information theory between spatially independent systems has been lost in the
setting of QFT. The restriction of the vacuum state to a local algebra
$\mathcal{A(O})$ is not a vacuum of the smaller region as it would be in QM
and as one might naively expect in QFT. Rather such a reduced vacuum is a
so-called KMS state i. e. a thermal state with respect to a modular
Hamiltonian which is intrinsically associated to the localized algebra
together with the vacuum. This phenomenon was first perceived in the context
of localization behind event horizons in the case of black holes and the Unruh
Gedankenexperiments. In that case the modular Hamiltonian has a physical
significance in terms of Killing symmetries and the thermal aspect accounts
for the Hawking radiation and the Unruh Gedankenexperiment in which the local
algebra is the wedge-localized algebra $\mathcal{A}(W)$ with is causal upper
horizon $H(W).$

Thanks to the modular localization theory we now know that all these
observations are consequences of the monad structure\footnote{The perhaps
conceptual distance between QFT and QM arises in the much stronger relational
(holistic) characterization of QFT in terms of the modular positioning of a
finite number of monads in a joint Hilbert space \cite{interface}.} of the
local algebras \cite{interface}, the only type of algebra which is consistent
with causal localization and which leads to thermal manifestation from modular
localization. A characteristic feature of this algebraic structure is the fact
that sharp localization causes infinite vacuum polarization at the
localization boundaries which in turn leads to localization-energy or
localization-entropy with respect to the vacuum to be infinite instead of zero
as in QM. The quantum mechanical analog is a global KMS state obtained from
thermal Gibbs states quantized in a box of volume V in the thermodynamic limit
V$\rightarrow\infty$; such states correspond to an heat bath in an open system
and are not related to localization. The algebra changes its nature in this
limit and the box quantized type I standard quantum mechanical algebras passes
into a limiting "monad" with a volume diverging energy/entropy \cite{Haag}.
There exists an analog to the thermodynamic limit in the local QFT case: the
"split limit", whereby one approximates the local monad by a sequence of type
I factors which are localized inside the an augmented double cone
$\mathcal{O}_{\varepsilon}$ with $\varepsilon$ being the size of a collar
around $\mathcal{O}$ within which the vacuum polarization can attenuate; the
part in the collar is only "fuzzy" (not sharp) localized in turns out that the
monad algebra $\mathcal{A(O})$ together with the collor results in a quantum
mechanical type I$_{\infty}~$factor. The close relation between the
thermodynamic volume divergence and that of the split limit $\varepsilon
\rightarrow0$ actually amount to much more than an analogy \cite{BMS}.

The contribution by Jordan to the Dreimaennerarbeit which is rightfully
heralded as the beginning of QFT would even nowadays serve as the perfect
example to illustrate all the points above which separate QFT from QM and
which have served as an never ending source of new conceptual ideas. A modern
discussion would start from the system of algebras generated by a current as
in (\ref{chiral}). A restriction of the vacuum state to a finite interval
leads to a singular KMS state with infinite entropy/energy.

In fact it was Heisenberg who pointed out this vacuum polarization caused
divergence problem to Jordan in 1931 (see \cite{Du-Ja}) before he wrote his
famous paper about boundary-caused vacuum polarization in partial charges
\cite{Hei}. Nowadays we know that these warnings against retaining only a few
frequencies (or occupying levels in the spirit of QM) were well-founded
because QFT has a \textit{holostic} structure. Cases where such simple
arguments go completely wrong are known in connection with simple minded level
occupation methods in order to compute the cosmological constant (see also
\cite{Ho-Wa}). Here \textit{holistic} implies that one cannot understand the
infinite degrees of freedom of a quantum field theoretic system by separating
it into finite sets. Jordan's problem of the energy fluctuation in a finite
interval of a chiral theory localized on the lightray is particularly suited
to be treated with the intrinsic and holistic modular localization methods of
QFT; in chiral models the leading coefficient of the localization
entropy\footnote{The dimensionless entropy is the simplest quantity which
shows the thermal behavior of localization onto a subinterval.} in
$\varepsilon\rightarrow0$ can be calculated exactly. The thermal nature of the
interval-restricted vacuum would have been of great conceptual and
calculational help, but not having these (intrinsic to QFT) concepts available
at this early time, Jordan used quantum mechanical methods which consisted in
performing a frequency decomposition of the field $u(t,x)$ and treating the
resulting system of infinitely many oscillators as a problem of QM with
infinite degrees of freedom\footnote{One note of caution, the use of "string"
in the old papers has nothing to do with string-localized fields since the
underlying field in Jordan's model is pointlike. The "string" refers to the
linerar version of a box-quantization.}. A treatment in the same quantum
mechanical spirit but with Jordan's tacit assumptions being spelled out more
explicitly is contained in \cite{Du-Ja}.

Problems as divergencies caused by sharp localization, which only were
understood recently, had to be dealt with in a somewhat artistic manner in
Jordan's time. In fact the reason for a certain critical distance which his
coauthors Born and Heisenberg kept to his more revolutionary colleague was
related to some ad hoc assumptions and artistic jumps in Jordan's calculation
which resulted from the fact that short of a conceptual framework for how to
deal with problems of QFT one had to rely on less than perfect quantum
mechanical methods. It would be a misinterpretation of the situation to view
Born and Heisenberg as being too conservative with respect to their
brainstorming young colleague; in contrast to Heisenberg's discovery of QM
less than a year before, Jordan's QFT did not come with ready to use
conceptually supported computational tools. In fact it took more than 80 years
of conceptual research to find the present setting.

Even at the risk of making statements which sound exaggerated, it is just the
fact that the quantum mechanical approach in Jordan's work and in its recent
improved review \cite{Du-Ja} did not solve the \textit{thermal side of the
conundrum} which is the strongest indicator that \textit{Jordan did not just
discover a another model of QM with infinite degrees of freedom but rather
came across a completely new conceptual setting}. The insufficiency of the
quantum mechanical operator techniques and the birth of QFT are two different
sides of the same coin and the critique of his colleagues, in particular of
Heisenberg, who a few years later discovered the quantum field theoretical
vacuum polarization, was well aimed.

The existence a strong critical counterweight against a new speculative idea
is the best indicator of a healthy particle theory. It is the disappearance of
this counterweight and the demise of the "Streitkultur" since the 80s when
those, who by their reputation and articulation were the natural candidates to
fill this role became instead the propagandists and salesmen of new
speculative ideas that led to a conceptual deterioration and a schism within
particle theory.

Unlike QM where its discovery almost immediately led to various computational
settings within mathematical control as well as a rapid understanding of its
underlying philosophy, this process took a much longer time in QFT owing to
its extraordinary subtlety and richness. The great perturbative progress in
the post world war II renormalization theory ended with impressive numerical
successes, but there was also the sobering conclusion that the diverging
perturbative series has at best the status of asymptotical convergence for
infinitesimally small coupling. This is certainly not sufficient in order to
secure the existence of a model. Only very recently the setting of modular
localization, an extremely different construction far removed from Lagrangian
quantization and functional integrals \cite{Sch}\cite{AOP}\cite{Lech}, led to
the existence of a certain class of two-dimensional models with a nontrivial
factorizing S-matrix. It is interesting to note that these methods use the
holistic modular localization ideas in an essential way i.e. the same concepts
which are important in the complete solution of the Jordan-Einstein
fluctuation conundrum \cite{interface}\cite{BMS}.

Relativistic causality entered Jordan's work for the first time explicitly in
the Jordan-Pauli calculation \cite{J-P} of the photon two-point function and
played an important role in his 1929 review of the first phase of QFT
\cite{Kharkov}. But it would be far-fetched to conclude from these occasional
flare ups of new ideas that causal locality was clearly seen as the
characteristic feature of QFT that sets apart QFT from QM. What prevented the
formation of such a viewpoint for a very long time was the fact that
independent of their differences, QM and QFT shared the Lagrangian and the
functional integral setting which led to overlooking the significant
conceptual differences. Up to the present it is not uncommon to find the
misleading terminology "relativistic quantum mechanics" instead of QFT; As
shown in section 2, interacting relativistic QM is something else than QFT. It
is precisely the result which causal localization has on the mutual coupling
of all states with the same superselected quantum numbers which led us to
refere to this characteristic property as "Murphy's law" and the related
property "nuclear democracy". In QM on the other hand, one can freely couple
or decouple channels as one wishes.

This extraordinary holistic nature of QFT is precisely what renders it more
fundamental than QM. But this comes at a prize, since it also makes QFT less
susceptible to quantum mechanical computational techniques. Single operator
methods of QM as, the construction of a Hamiltonian in terms of fields lead to
well-defined renormalized expectation values only through ill-defined infinite
intermediate steps (cutting off integrals, introducing ad hoc regulators in
order to enforce finiteness of integrals) and there is no guaranty that these
manipulation preserve the Hilbert space structure of the theory. The intrinsic
method which is solely based on localization\footnote{The most prominent
renormalization theory of this kind is the Epstein-Glaser setting.} and a
certain minimality principle has no infinity but in certain cases requires to
introduce additional couplings which was not explicitly there at the beginning
(or to nonrenormalizable theories with infinitely many parameters). The modern
era of QFT as a setting in QT which comes with its own set of concepts
different from those of QM started at the end of the 50s with a programmatic
talk by Rudolf Haag \cite{Lille}.

As explained in a previous subsection the foundational distinction of the two
quantum theories is localization. Jordan, who after Einstein and Heisenberg
was one of the most conceptual-philosophic motivated among 20th century
physicists\footnote{Among the early books on QFT his "anschauliche
Quantentheorie"\cite{anschau} is certainly the only one which dedicates entire
chapers to the problem of finding a new conceptual-philosophical setting for
interpreting the new theory.}, left QFT shortly after the phenomenon of vacuum
polarization was noticed in the context of "partial charge" (charge localized
in a sphere \cite{interface}) in free field theories by Heisenberg and in
interacting theories by Furry and Oppenheimer who noted that an interacting
field applied to the vacuum create a one particle state accompanied by an
infinite vacuum polarization cloud consisting of particle-antiparticle pair
states. These were the first indications that there was a dramatic difference
to QM. It is precisely these interaction-caused polarization clouds which
determine the ultraviolet behavior of QFT and which, if not properly treated
lead to ultraviolet divergences dealt with in renormalization theory. We know
nowadays that localization causes thermal behavior including "localization
entropy" \cite{interface}\cite{foun}. It also causes all operators to
communicate with all states having the same superselected charges; the
Murphy's law leading to "nuclear democracy" without which Jordan's idea of a
d=1+3 photon theory of neutrinos cannot even be formulated.

We remind the reader that the Einstein-Jordan conundrum from section 1 has
reappeared here because it shares with the bosonization/fermioniszation issue
the same QFT formalism of chiral currents (\ref{chiral}). But whereas the
fluctuation conundrum has (less simple) extensions to higher no-conformal
spacetime models, there is, as mentioned before no conceptual support from QFT
for Jordan's use of two-dimensional bosonization/fermionization model as a
realistic neutrino theory of light. Whereas typical quantum mechanical objects
as oscillators can be adjusted to arbitrary dimensions (which justifies a
pedagogical presentation in one spatial dimension), this is not the case in
QFT. A Wigner particle as the photon and its free field is an object in d=1+3;
Wigner's representation theory of the Poincar\'{e} group and the free QFT
theory following from it in a functorial way depends on d=1+3, only there the
characterization in terms wave functions and covariant Maxwell fields
describes photons. But if one wants to criticize Jordan on this point, one has
to include all his contemporaries (including his deriders) since all of them
believed that a check of a structural property in the context of lowest
possible dimension is enough for the validity in higher dimensions. Indeed
none of them criticized him on dimensional grounds, rather those who
articulated themselves as Fock \cite{Fock} thought that he made a
computational mistake\footnote{Not so uncommon, before I red Klaiber
\cite{Klaiber} I also sensed that there was something wrong with this
two-dimensional formalism.}.

There is an interesting connection of the discrete "Paulion" formalism which
appears in the famous Jordan-Wigner paper \cite{J-W} and the fermionization
(\ref{fer}) which for generic charge $\alpha$ is really an "anyonization". The
Jordan-Wigner transformation formalism becomes more concrete in d=1+1 when the
abstract ordering passes to a concrete linear ordering. In the continuous
limit one obtains exponentials of line integrals which make the nonlocal
character of the "anyonic" (braid group) commutation relation explicit
(related to the "fermionization"). Jordan has various footnotes \cite{Jo} to
the Jordan-Wigner paper but he dids not elaborate the connection between the
two formalisms.

Jordan's model (\ref{fer}) has an interesting relation with the later
Schwinger model \cite{Schwinger}. The latter is a 2-dimensional massless
quantum electrodynamics which Schwinger proposed in order to illustrate that a
gauge theory is not necessarily describing photons and free electric charges,
rather its observable content under certain circumstances may consist of
massive vectormesons and screened electric charges free of any charge
selection rule. In fact the gauge invariant content of the Schwinger model is
described by a field which is the exponential of a massive free field i.e. it
is of the form (\ref{fer}) except the field $\Phi_{schw}$ is now a
massive\footnote{The mass is actually proportional to the square of the
coupling strength.} free field in d=1+1 which depends on space and time and
not just on the lightray combination u=t+x. The model looses its screening
aspect for short distances when%
\begin{equation}
e^{i\alpha\Phi_{schw}(x,t)}\overset{s.d.}{\rightarrow}e^{i\alpha(\Phi
(u)+\Phi(v))},~u=t+x,~v=t-x
\end{equation}
The short distance limit is carried out on the n-point correlation functions
by scaling the fields with the right mass power in such a way that no
correlation diverges. Then there are many n-point function which go to zero
namely all those which violate the charge superselection rule (\ref{cor});
this is how the charge superselection emerges. Without this prescription the
correlation functions would not belong to a Hilbert space. It turns out that
the short distance limit of the massive Schwinger model is precisely the
massless Jordan model which with a lot of imagination one could see as charge
liberation (the Jordan model) in the short distance limit of the charge
screened QED$_{2}$ (the Schwinger model).

This kind of screening is the two dimensional analog of the Schwinger-Higgs
screening. The situation changes in case of n-component \textit{massive}
fermions; it has been suggested that these models describe the two-dimensional
analog \cite{BSRS} of the still unsolved problem of quark confinement in
4-dimensional quantum chromodynamics (QCD). The liberation of charges in the
asymptotic short distance region of the Schwinger model corresponds to the
perturbative established asymptotic freedom in the nonabelian gauge
description of QCD. Such a simple illustration of screening/confinement versus
short distance charge liberation is only possible in d=1+1; free massless
fields in higher dimensions do not permit such constructions. Hence although
the Jordan model, different from the intentions of its protagonist, has no
bearing on a neutrino theory of light (for whose validity there is not the
slightest hint within the Standard Model, which is our presently best particle
theory), it is believed to serve as a useful analogy for important unsolved
problem of quantum chromodynamics which is the conundrum of quark confinement.
A recent description of the Jordan model and its appearance in the massless
limit of the Schwinger model can be found in \cite{2-dim}.

The fermionization formula (\ref{fer}) of the Jordan model describes a chiral
fermion only for one value of $\alpha$ whose square $\alpha^{2}$ corresponds
(in a suitable normalization of the current) to the operator dimension
dim$\Psi=1/2\footnote{The normalizations used in Jordan's work is different
from that used in more recent times. We found it convenient to state the
results without fixing normalizations.}.$ For this value the $\Psi$ can be
written as a Fourier transform in terms of Wigner particle creation and
annihilation operators $a^\ast(p),a(p)$ and their charge-conjugate
antiparticles. For other values of $\alpha$ one encounters a fields whose
commutation relations are those of an anyon i.e. an object with commutation
relations which go beyond the Boson/Fermion alternative and are related to a
representation of the braid group. Such fields were proposed as a model of an
"infraparticle" in the beginning 60s \cite{infraparticle} which is a
particle-like object which is behind the breakdown of standard scattering
theory (the Bloch-Nordsiek infrared divergence problem). The reason is that
for those charge values $\alpha\neq$ semiinteger the pointlike field
description was lost in favor of an infinite stringlike behavior as in
(\ref{fer}). The model on which this was first noticed, was a two-dimensional
massive Dirac field coupled to the derivative of a massless scalar field. The
long range interaction leading to a solution of the form%
\begin{equation}
\psi(x)=\psi(x)_{0}e^{ia\varphi(x)}=\psi(x)_{0}e^{ia\int_{-\infty}^{x}%
\partial_{\mu}\varphi(x)dx^{\mu}}\label{infra}%
\end{equation}
Here $\psi_0$ is massive free Dirac field and since the infrared divergent
zero mass scalar field is better interpreted as a string localized free field
we prefer the second representation. The momentum space manifestation of the
stringlike localization shows an interesting modification at the mass shell
where in all theories describing particles there would be a $\delta
(\kappa^2-m^2)$ function. In terms of the Kallen-Lehmann spectral function the
infraparticle spectral density for a massive d=1+1 infraparticle is instead%
\begin{align}
\left\langle \psi(x)\overline{\psi(y)}\right\rangle  &  =\frac{1}{2\pi}\int
d\kappa^{2}\rho(\kappa^{2})\int\frac{dp}{2\sqrt{p^{2}+\kappa^{2}}}p^{\mu
}\gamma_{\mu}e^{ip(x-y)}\label{KL}\\
\rho(\kappa^{2}) &  =c(\alpha)\theta(\kappa^{2}-m^{2})(\kappa^{2}%
-m^{2})^{-d(a)}\nonumber
\end{align}
The only important aspect of this formula is that a) for vanishing coupling
strength $\alpha$ the infraparticle power law of the spectral density passes
into the free field result $\rho(\kappa^2)=\delta(\kappa^2-m^2)$ and b) it is
not possible to write the spectral function as $\rho(\kappa^2)=\delta
(\kappa^2-m^2)+rest$ without violating the positivity of the rest i.e. In an
infraparticle Hilbert space there is no possibility to sneak in a particle
through the back door.

The infrared behavior of the perturbative treatment of this model are
reminiscent of those encountered in QED which led to the strong suspicion that
the minimal interaction of electron/positrons with photons is so strong in the
infrared (for long distances) that the structure of the particle itself is
affected i.e. the changes are more radical (see next section) than in the
Coulomb interaction in QM where the modification of time dependent scattering
theory leaves the structure of particle unaffected.

The study of chiral conformal QFT started in the beginning of the 70s with the
Jordan model but without the knowledge about Jordan's work since the title
under which he published it did not sound trustworthy. All the global
conformal block decomposition results involving the universal conformal
covering representation were checked for the Jordan model. Only with the
appearance of the family of minimal models by Belavin, Polyakov and
Zamolodchikov it became clear that the prior structural work was not in vain,
the expected richness of chiral conformal QFT really materialized.

By now the wealth of very nontrivial results is staggering, the world of
chiral models is meanwhile under impressive mathematical control. This and the
infinite family of massive factorizing models are the only families where
existence proofs of models have been achieved. It seems that Jordan had the
right conceptual-mathematical instinct in emphasizing these models in many of
his publication.

The multi-component extension of the Jordan model%
\begin{align}
j_{k}(u)  &  =:\psi_{k}^{\ast}(u)\psi_{k}(u):=\partial\Phi_{k}(u),~k=1,...n\\
\Psi(u,\vec{\alpha})  &  =e^{i\sum_{k=1}^{n}\alpha_{k}\Phi_{k}(u)}\nonumber
\end{align}
is the starting point of a rather simple family of models whose maximal local
extensions are characterized by even lattices. Their representations are
classified by the finite number of dual lattices and there is a finite number
of selfdual lattices which are connected with finite exceptional groups, the
largest being the so-called \textit{moonshine group}.

Another application in which one quantizes $\Phi_{k}$ with a n-dimensional
zero mode attached to a n-dimensional quantum mechanics $p_{k},q_{k}~k=1,..n$
was used for the operator formulation of the dual resonance model which led to
string theory.

The history of the Jordan model in conjunction with the Schwinger model
reveals that conceptually rich models develop a life of their own and are even
able to survive flawed reasons which served their original introduction. One
can be sure that at the time of the \textquotedblleft neutrino theory of
light\textquotedblright\ mock song neither Jordan nor his satyric colleagues
had any firm idea about what message this 2-dim. model was supposed to reveal.
In the 30's the extraordinary subtle conceptual and mathematical problems
posed by interacting QFT was not yet appreciated and the idea of studying
soluble models as a kind of theoretical laboratory in order to learn something
about the classification and construction of interacting particles was still
in the distant future. After Jordan's series of papers on the neutrino theory
of light there were several other authors who published papers under the
heading of neutrino theory of light without even mentioning how this can
achieved in the realistic d=1+3 case of physical photons and neutrinos. There
were several authors who continued to publish articles on this 2-dim. model
under the title \textquotedblleft neutrino theory of light\textquotedblright%
\ without bothering how to get to light and neutrinos in d=1+3 QFT world. It
remains somewhat incomprehensible why in none of these papers commented on the
discrepancy between title and content.

\section{Nonlocal gauge invariants and an algebraic monopole quantization}

Jordan was the first who realized that the passing from the classical
electrodynamic to its quantum counterpart brought about a loss of locality.
More specifically besides the local observables which in the gauge theoretical
setting are by definition the quantum counterparts of the classical (second
kind) gauge invariants any physical charge-carrying object cannot be better
localized than a spacelike semiinfinite string. That the algebra of all
physical fields is larger than that generated by local observables is not
exceptional, but that electric charge-carrying operators cannot be compactly
localized is unusual in a theory which has the name "local" is unusual. It
raises the question which structure is responsible that beyond charge neutral
local observables the charge sectors of a theory are nonlocal in that strong
sense. It can be shown that any renormalizable theory which a charged current
is related to zero mass s=1 field strength through a Maxwell equation as in
QED falls into this class.

The best localization for a charged generating field is that of a semiinfinite
Dirac-Jordan-Mandelstam string (DJM) characterized \textit{formally} by the
well-known expression%
\begin{align}
\Psi(x;e)  &  =~"\psi(x)e^{\int_{0}^{\infty}ie_{el}A^{\mu}(x+\lambda e)e_{\mu
}d\lambda}"\label{DJM}\\
\Phi(x,y;e)  &  ="\psi(x)e^{\int_{0}^{1}ie_{el}A^{\mu}(x+\lambda
(x-y))(x-y)_{\mu}d\lambda}\bar{\psi}(y)" \label{br}%
\end{align}
Such objects involving line integrals over vectorpotentials or their bilocal
counterparts with a connecting "gauge bridge" (\ref{br}) appear already in
Jordan's work on attempts to quantize in a gauge invariant manner
\cite{bridge}. As an ardent positivist and a fierce defender of Heisenberg's
maxim to use observables throughout, he even tried to formulate the dynamics
of quantum electrodynamics solely in terms of gauge invariant operators. As a
similar attempt more than 20 years later by Stanley Mandelstam, the effort
fell short of what the authors had expected.

There is however one very pretty fall-out of these attempts which is worth
mentioning. Being impressed by Dirac's magnetic monopole quantization, but not
by his too classically looking geometric method of presentation, Jordan
published a very different algebraic operator derivation in the same year
\cite{Mono}. Three years later he returned to this topic, this time presenting
his arguments with more details and some helpful drawings. The argument in
both papers is based on the use of the above "bridged" bilocals $\Phi
(x,x^{\prime}).$ Starting from the commutation relations%
\begin{align}
&  \left[  \Phi(x,x^{\prime}),\Phi(y,y^{\prime})\right]  =\delta(y-x^{\prime
})\Phi(x,y^{\prime})e^{i\omega(x,y^{\prime},x^{\prime})}-\delta(x-y^{\prime
})\Phi(y,x^{\prime})e^{i\omega(y,x^{\prime},x)}\\
&  \omega(x,y,z)=magnetic~flux\text{ }through\text{ }triangle\nonumber\\
&  Jacobi\text{ }identiy\text{ }for~\Phi^{\prime}s\curvearrowright
e^{i\sum_{tetraheder}\omega}=1\nonumber
\end{align}
The last line denotes the application of the Jacobi identity to the bridged
bilocals whose validity turns out to be equivalent to the magnetic flux
through a tetrahedron being integer valued (in certain unities). A similar
method for monopole quantization where the result also emerges from
\ cohomological argument involving a tetrahedron has been proposed by Roman
Jackiw \cite{Jack}.

The quotation marks in the above formulas (\ref{DJM}) highlights their formal
aspects. Since these objects are not belonging to the local gauge invariant
operators which appear in the formalism of n$^{th}$ order renormalized
perturbation theory, they have to be defined (and renormalized) by hand, a
gruesome task which was carried out by Steinmann \cite{Stein} who succeeded to
attribute mathematical meaning to these expressions. This technical work is
important because electrically charged fields cannot be better localized than
along a semiinfinite spacelike string and this has radical implication for the
associated charged particles.

It is fascinating and very informative to follow the idea of gauge invariant
nonlocal semiinfinite string-localized fields and that of bridged bilocals a
bit more through the history of particle physics. In an historically important
paper, written about the same time as the appearance of the above
string-localized fields in Jordan's work, Bloch and Nordsiek, using a simple
model, argued that the scattering of photons off charged particles will lead
to infrared divergencies unless one treats the problem in the way they
proposed in their model. After the renormalization theory of QED was
understood, Yennie, Frautschi and Suura (YFS) showed that although the
scattering amplitudes are infrared divergent the inclusive cross section for a
specified photon resolution $\Delta$ not; the artificially introduced cutoff
from the "virtual" photons in the Feynman amplitude compensates with that
coming from the integration over the infrared tail of the real photons to be
summed over up to $\Delta.$

From a pragmatic view the formalism of YFS (i.e. compensating two infrared
divergencies against each other) would have been the end; a finite answer
which agrees with the experimental data would signal for many physicists:
mission accomplished. But Jordan and some of his contemporaries had a strong
philosophical motivation and one can almost be sure that this kind of
pragmatic reasoning would not have been the end of the infrared issue in QED.
Rather it would have been very much in the spirit of Jordan to search for a
deep connection between the formula for the string-localized generating fields
describing charge-carrying fields (\ref{DJM}) and charged "particles". The
parenthesis is to indicate that electrically charged particles in QED are not
particles in the usual (Wigner) sense of irreducible (m,s) representations of
the Poincar\'{e} group\footnote{The relativistic particle concept was laid
down in Wigner's famous representation theoretical classification of 1939
\cite{Wig}.}. In this way it becomes clear that the breakdown of the
scattering theory is not just because the interaction is long ranged as the
quantum mechanical Coulomb scattering where the breakdown of the standard
scattering theory through the appaerance of a logarithmic phase factor does
not have any consequences for the structure of single particle states. Rather
in QED the very existence of particles is affected, electrically charged
particles are "infraparticles" and even in case of a one particle state the
best one can do is prepare such a state that no photon with energy larger than
$\Delta$ can emerge from such a state where $\Delta$ can be arbitrarily small.

In the previous section the issue of infraparticles came up in connection with
a hidden infinite stringlike localization aspect of Jordan's two-dimensional
model. In that case the two-point spectral function $\rho(\kappa^{2})$ remains
covariant. However in 4 dimensions the spacelike string direction defined in
terms of a spacelike unit vector $e$ certainly enters the covariance low for
$\rho.$ The characteristic dissolution of the mass shell delta function into a
singular cut, which starts at $p^{2}=m^{2}$ is a general feature of
infraparticles. Unitarity puts a lid on the strength of the singularity, it
must be milder than a delta function. This leads to a vanishing in/out LSZ
scattering limit; the perturbative infrared divergence of the scattering
amplitude is a perturbative phenomenon; by summing up the leading terms and
letting $\Delta\rightarrow0,$ the infinity is converted into zero.

Hence the pointlike gauge dependent Dirac spinor $\psi(x)$ which enters the
Lagrangian is an unphysical chimera which has to be tolerated as an
intermediate computational tool as long as one does not know how to formulate
the dynamics and the computations directly in terms of physical operators.
There is simply no compact localized operator which applied to the vacuum
generates a state with an electron or positron charge, only zero charge
operators are compactly localizable. This is an illustration of the fact that
the main and only physical principle is causal localization and its
realization in Hilbert space (unitarity). Even the Lagrangian formalism and
perturbation theory has to cede if it produces unphysical operators. The only
known way up to know is to repair the perturbative situation by hand i.e. to
construct the DJM object from the unphysical $\psi$ and $A_{\mu}$ which is a
gruesome enterprise. The fact that QFT is governed by one principle, namely
causal localization, does not make life simpler.

There is no doubt that the knowledge of the local observables in principle
fixes the remaining physical operators, including bridged bilocals. But the
question which remained unanswered since the time of Jordan is how, by what
formula? If the theory would not be a gauge theory but one which only involves
s=0 and s=1/2 fields it would be less problematic to construct bilocals from
locals. The bilocal $:A(x)A(y):$ can be obtained from the product of two
locals $:A^{2}(x):$ by a lightlike limiting procedure \cite{L-S} and recently
the problem whether bridged bilocals can be obtained in this way came up
\cite{Jacobs}. Having such zero charge bilocals, the DJM would be expected to
result from the "shifting the (unwanted) charge behind the moon" argument.

There is however another more radical (but also more promising) method. The
origin of all the problems goes back to Wigner's 1939 classification of
positive energy representations of the Poincar\'{e} group. The transition from
Wigner's form to the covariant form of the representation is not unique. In
the massive case the (m,s) s=halfinteger Wigner representation can be
described by infinitely many dotted/undotted spinorial fields for given spin s%
\begin{align}
&  \psi^{(A,\dot{B})}(x),~\left\vert A-\dot{B}\right\vert \leq s\leq A+\dot
{B}\label{spin}\\
&  m=0,~s=\left\vert A-\dot{B}\right\vert \nonumber
\end{align}
The second line contains the formula for the significantly reduced number of
zero mass spinorial wave functions; the popular potentials $A_{\mu}$ for s=1
and $g_{\mu\nu}$ for s=2 are not backed up by representation theory which
would allow field strength ($F_{\mu\nu}$ for s=1, $R_{\mu\nu\kappa\lambda}$for
s=2). In classical physics there is no such requirement because unitarity and
Hilbert space requirement is no issue. If one uses pointlike covariant
potentials for the formulation of interactions (the standard method in QED)
then one has to rely on a quantum gauge formalism (Gupta-Bleuler or BRST)
which allows a cohomological return to a physical subspace which does not
contain charged particles.

But there is another way which is more physical. Its starting point is the
realization that the full spinorial possibilities (\ref{spin}) can be
recovered with semiinfinite string-localized covariant potentials $A_{\mu
}(x,e),g_{\mu\nu}(x,e)$ $e~$= spacelike string direction. In this way the
origin of the string-localized charged fields and their infraparticle behavior
near the old mass shell are not surprising since the semiinfinite localization
aspect enters through the vectorpotentials from the beginning. Although all
the fields are now living in a physical Hilbert space, the Epstein-Glaser
iteration step is more complicated since the causal position of semiinfinite
strings, which should preserve the string localization for counterterms, is
now more involved. These new ideas may lead to the long overdue reformulation
of gauge theory.

Closely connected with the string localization of the potentials is the
Aharonov-Bohm effect. In fact there is a theorem which shows that behind this
effect is the consequence of some basic structural difference between operator
algebras generated by field strength associated to massless spin $s\geq1$ as
compared to their massive counterpart. Whereas for any compact simply
connected spacetime region $\mathcal{O}$ there holds Haag duality%
\begin{equation}
\mathcal{A(O})^{\prime}=\mathcal{A(O}^{\prime})
\end{equation}
i.e. the commutant of the algebra of $\mathcal{O}$-localized operators equals
the algebra of operators localized in the causal complement of $\mathcal{O}$,
if it comes to non simply connected regions (example toroidal spacetime
regions $\mathcal{T}$) the $m=0,s\geq0,$ the operator algebras show a
violation \cite{Roberts}%
\begin{equation}
\mathcal{A(T})^{\prime}\supset\mathcal{A(T}^{\prime})
\end{equation}
of Haag duality which has no counterpart in the massive case. The explanation
is that operator algebra $\mathcal{A(T})$ generated by the field strength does
not contain all physical operators, there are Bohm-Aharonov like nonlocal
operators in $\mathcal{T}$ , but thanks to the breakdown of Haag duality the
algebra of the field strengths "knows about its own imperfection". For higher
spin the Bohm-Aharonov phenomenon has a generalization affecting also algebras
localized in multifold connected regions. The use of stringlocalized
potentials $A_{\mu}(x,e),~\ g_{\mu\nu}(x,e),...$makes this nonlocal aspect
manifest \cite{MSY} which remains hidden in the standard formalism and only
suddenly pops out e.g. when one tries to construct electrically charged
operators assuming that by some extension of perturbation theory one can get
to such objects.

It is quite surprising that a theory as QED, which already in the middle of
the 30s showed a rich conceptual structure, has still not reached its
conceptual closure. This is even more so for its nonabelian extension the Yang
Mills theory and quantum chromodynamics (QCD). We have gotten accustomed to
nice words as "gluon" and "quark" confinement about which we think we know
their content, but our understanding goes hardly\footnote{Whereas the vacuum
correlations of gauge-invariant local observables are finite in the abelian
case (and only on-shell formfactors and scattering amplitudes are infrared
divergent ) in the nonabelian case also the gauge invariant correlation
functions are divergent.} beyond the small subset of gauge invariant local
operators. When it comes to the description of physical charge-carrying
operators our formalism forsakes us not to mention the string localized
counterparts of the DJM operators and the question what happened to the gluon
and quark degrees of freedom.

Whereas the long stagnation on questions like this and on the standard model
could be shrugged off by pointing to the complexity of the task, it is
somewhat saddening to notice the loss of hard gained conceptional
understanding. A typical illustration for the conceptual impoverishment is the
story of infraparticles which begun with the infrared divergent scattering
amplitudes and the recipe for calculating inclusive cross sections with a
given inclusive resolution $\Delta.$ As mentioned, the conceptual conquest of
this problem started with the realization that the root of the problem is not
merely a long range modification of scattering as in the quantum mechanics of
Coulomb scattering, but a radical change of the particle concept. From
Jordan's semiinfinite stringlike charged fields (\ref{DJM}) to infraparticles
is a long way. The first observation of a deviation from the standard particle
structure was observed in a two-dimensional model \cite{infraparticle} similar
to Jordan's in the previous section. It was observed that the expansion of the
cut which starts at the would be particle mass with respect to the coupling
strength leads to similar terms as in the YFS work. The infraparticle aspect
of electrically charged particles in QED was proven in the 80s, the most
conceptual line of arguments in \cite{Buch} established the infinite extension
of electrically charged infraparticles as a consequence of the quantum
adaptation of Gauss's law. At that point it became clear that the string-like
extension is inexorably linked to the "dissolution" of the mass shell.

There are some examples in the history of particle physics where, as if led by
an invisible hand, two separate discoveries which are different sides of the
same coin are made at the same time but their internal connection is only seen
many decades afterwards. Certainly the discovery of the infrared phenomenon in
QED by Bloch and Nordsiek \cite{Bl-No} and Jordan's semiinfinite spacelike
localization of the physical charge generating operator in QED belongs to
these cases. Being aware of this connection, it is not so surprising that the
oldest mathematically controlled 2-dimensional models of infraparticles where
constructed with the exponentials of zero mass field of the previous section,
since the zero mass field in d=1+1 is really semiinfinite string (in this case halfline)-localized.

The loss of conceptional understanding in contemporary attempts to go beyond
the standard (Wigner) particle concept becomes obvious in the present flood of
papers on "unparticles", which appears as the particle equivalent of the
German word "Unsinn". The authors owe an answer how their ill-defined objects
can be placed into the conceptual quite dense meshwork of string-localized
fields and its momentum space properties in terms of dissolved mass
shells\footnote{Apparently the unparticle followers believe that the infrared
divergencies in the scattering theory of electrically charged particles
represent (like the quantum mechanical Coulomb scattering) just a small
modification of scattering theory and not a radical change of the particle
concept.} Even in those few cases where they cite the infraparticle work, it
is clear that they do not understand its conceptual basis. They seem to think
that the YFS kind of infrared problem is similar to the infrared aspects of
Coulomb scattering where the one-particle remain those of standard particles.
There seems to be nobody of sufficient knowledge of QFT whom they would listen
to. The new generation of referees have the same background and are unable to
lift the state of arts to where it has been in the past. This makes the
question of the causes behind this derailment relevant, but this article about
backtracks to Pascual Jordan is not the right setting to enter a critical
analysis of the Zeitgeist.

\section{An end of foundational QFT before its scientific closure?}

The present work described Jordan's discovery of QFT in connection with a
problem posed by Einstein's pro photon argument based on his fluctuation
formula and his subsequent fruitful controversy with Dirac's particle-based
viewpoint. It also tries to shed some light on Jordan's somewhat "futuristic"
idea, formulated at the 1929 Kharkov conference with the hindsight of later
developments in QFT which vindicated his rejection of a classical parallelism,
or in his own words his search for an understanding "independent of classical crutches".

Such a conceptual setting was discovered 30 years later by Rudolf Haag at a
time when, as the result of world war II and for other sociological reasons,
the continuity of particle physics research was interrupted and a good part of
the early history of QFT was forgotten. Haag sketched his project in which
causal locality took the center stage for the first time at a conference in
Lille 1958 \cite{Lille}, the first more detailed account \cite{H-S} in which
the causality principle consisted of the spacelike Einstein causality and the
timelike causal shadow property (time slice property) extended by the closely
related positive spectrum requirement. This formulation keeps all the
important physical properties without using a quantization parallelism and is
even independent of the chosen field coordinatization of local observables.

By removing Jordan's "classical crutches", LQP moves farther away from
quasiclassical and perturbative approximations and turns more to toward issues
which are structural in nature, owing to the fact that for a theory based on
principles it is easier to draw structural conclusions than to classify and
construct individual models within these defining principles. Thanks to the
mathematically powerful formulation of causal locality in the form of modular
localization during the last two decades, the first constructive results about
the existence of nontrivial chiral models as well as existence proofs for some
factorizing models mentioned in section 3.2 have emerged. With this modest
success, LQP has entered a terrain which has been totally out of reach for
standard QFT. Although not a bad start, such nontrivial models are test cases
still belong to a "theoretical laboratory". Whereas two-dimensional models of
QM capture the essential points, two dimensional models of QFT do not allow to
make structural conclusions about realistic higher dimensional models.

As in real life, the frustration resulting from running against a wall in
trying to solve a crucial problem leads to the invention of distractions in
the form of palliatives, in the Lagrangian approach to QFT they took the form
of: "effective QFT" and "living with infinities", to name a few. The new
concept of modular localization can be viewed as retaking Jordan's attempt to
get away from the parallelism to classical theories inherent in quantization
and move towards an intrinsic description of QFT and approximations which
preserve its holistic localization structure.

This program could not have been pursued at the time of Jordan since the
mathematical-conceptual foundation in this first phase after discovery was too
weak. But the motivation to place QFT on a foundational course reappeared
independently three decades later in the Haag and LSZ setting, where the
subtle connection between fields and particles was studied and its origin in
causal localization was identified. The profound understanding of these issues
could however not prevent the occasional appearance of misunderstandings. An
example without lasting consequences was the publication of an article in the
prestigious Phys. Rev. Lett. suggesting that Fermi's arguments by which he
wanted to show that the local velocity\footnote{In a relativistic QM only the
effective velocity (analog of the acoustic velocity) whereas the local
velocity is infinite.} in QED is c (as in its classical Maxwell counterpart)
is incorrect \cite{Heger}. This claim was echoed in an article by Maddox in
Nature and picked up by the world press. Besides bringing short-lived fame for
having removed theoretical obstacles against the existence of time machines,
the net result of this affair was a strengthening of the modular localization
of QFT \cite{Bu-Yn}. This illustrates the positive side of committing a
conceptually interesting error and afterwards resolving it with even more
interesting subtle arguments.

In contrast the misunderstandings underlying the interpretation of the dual
model and string theory were of a different caliber. One reason why they has
been immune against critique is that it is not the proposal of an individual
but of physicists embedded in a rather uncritical community, the precursor of
the later globalized community. This is perhaps the most inappropriate social
form for doing foundational research since globalized communities cultivate
physical monocultures and have no critical breaks. Up to the beginning of the
80s most speculative ideas where put to critical tests usually by more
established and prominent researchers. In a highly speculative area as
particle theory critique is essential for keeping the balance. The golden
years of particle theory and QFT in the years 1950-1975 were also the years of
"Streitkultur" going with the names as Pauli, Lehmann, Jost, K\"{a}ll\'{e}n,
Landau and although in the new world the tradition is somewhat different, one
may add names as Oppenheimer, Feynman, Schwinger etc.

The string culture did away with this, instead of individuals or small groups
localized at one physics department there is now a large globalized community
of people who have a similar scientific background and dedicate their
knowledge to the promotion of the theory around which the community was
formed. Those members who by there seniority would have played a critical role
in earlier times, act now more like gurus i.e. if criticism (as that about the
localization in string theory) comes up they will refudiate it and make sure
that no doubts linger on. To keep morals high, they praise the string theory
as the "gift of the 21st century to the 20th", "the only game in town" and
similar pronouncements (whereas critical individuals outside the community
think of it more as a undisposed relic which the 20th left to the 21th
century). Similar arguments which led to the claim that the quantization of
the Nambu-Goto Lagrangian describes a spacetime string brought the community
to accept the claim that the superstring contains quantum gravity and hence is
the millennium theory of everything (TOE), thus insuring its hegemony over the
future of particle physics.

Actually these tendencies are in fact attributes of the Zeitgeist and not
indicators of weakening intellectual capabilities. Ignoring critique and
warnings, in the belief to be in the possession of a "theory of everything" is
not different from the human faults which led to the crisis of the finance
capitalism. These developments do not accidentally occur at the same time,
rather they are connected by a strong sociological undercurrent.

This changed social environment is particular strongly felt in Germany, the
country where almost 90 years ago QM and QFT\ began and where a successful
balance between speculative innovations and their critical review was
maintained for a long time. This continuity was interrupted when more recent
times the strange idea gained ground that foundational research can be bred in
well-financed exellency centers by coupling the salary to the showroom value
of a person. Such a coupling of the scientific value and career of an
individual to the uncritical fashions cultivated in a globalized community is
detrimental for the course of science. Even though string theory has no
credible relation to the experimental reality of high energy physics
laboratories, the presence of string theorists is deemed necessary for
maintaining the laboratories international reputation. It is clear that
according to such criteria string theory is in a superior position, especially
since its erronous handling of localization in relativistic QT occured on such
a subtle issue of particle theory which is far beyond textbook QFT and out of
reach for most physicists. Conceptual errors in times of monocultures of
globalized communities without critical breaks are not like those errors
commited by individuals in the time of Jordan (of which some were mentioned in
this article); they were usually rapidly understood and sometimes their
correction led to a more profound understanding as compared to a direct
progress without the erronous sidestep. Pauli's famous "not even wrong" has to
be understood in this context.

It is therefore not surprising that after Berlin in the 90s, Munich (the MPI)
at the beginning of this decade, now also Goettingen will not have QFT in its
Theoretical Physics department. The expected loss of its academic basis at the
university of Hamburg will set the seal on a more than 80 year tradition of
innovative QFT in Germany. The University of\ Hamburg was for a long time the
home of the renaissance of QFT after world war II (Lehmann, Haag); even though
one chair went to string theory it still maintained its high level of quantum
field theoretic research which it already had right after its foundations with
Pauli's presence in the 20s\footnote{Wolfgang Pauli as well as his senior
Wilhelm Lenz, were products of Sommerfeld's creative hotbed of the new quantum
physics in Munich.}; in fact his famous epiphany which led to the exclusion
principle, occurred in Hamburg, next to G\"{o}ttingen one of the great centers
of the historical quantum dialogue at the beginning of QT.

Many particle theorists hope that the results of the LHC experiments will help
to get out of the present already more than 30 years lasting stagnation. It is
difficult to imagine how this can happen in a situation where some of the most
fashionable theoretical ideas (those leading to extra dimensions) originated
from misunderstandings. Experiments cannot correct flawed theoretical ideas,
and the experimental falsification of an incorrect theory is epistemologically
void. the role of experiments is to exclude mathematially and conceptually
correct models and to narrow the new search by continuing this process with
respect to the reduced set of remaining models.

With the discontinuation of foundational QFT research groups around chairs at
theoretical physics departments of German universities in the land in which
Pascual Jordan discovered QFT in 1925 and which succeeded to recover from the
intellectual losses of its Nazi past after world war II and again became a
leading place for foundational innovations in QFT, this role now seems to come
to an end. This occurs at a time when new ideas promise to revolutionize this
theory a third time. Research on QFT requires foundational knowledge and a
long breath and therefore depends on on continuity as no other area in
physics. Once interrupted for one generation it probably cannot be re-created;
as a result the foundational research in QFT may end (unlike that in QM which
is a foundationally complete theory) before it reached its conceptual closure.

\end{document}